\documentclass[12pt]{article}
\usepackage{hyperref}
\usepackage{cite}
\usepackage{color}
\usepackage{graphicx}
\usepackage{amsmath}
\usepackage{amssymb}
\usepackage{xspace}

\makeatletter
\@addtoreset{equation}{section}

\makeatletter
\renewcommand\section{\@startsection {section}{1}{\z@}%
                                   {-3.5ex \@plus -1ex \@minus -.2ex}%nn
                                   {2.3ex \@plus.2ex}%
                                   {\normalfont\large\bfseries}}
\renewcommand\subsection{\@startsection{subsection}{2}{\z@}%
                                     {-3.25ex\@plus -1ex \@minus -.2ex}%
                                     {1.5ex \@plus .2ex}%
                                     {\normalfont\bfseries}}

\def\baselinestretch{1.2}
\newcommand{\be}{\begin{equation}}
\newcommand{\ee}{\end{equation}}
\newcommand{\beq}{\begin{eqnarray}}
\newcommand{\eeq}{\end{eqnarray}}

\newcommand{\gone}[1]{{}}

\begin{document}
\begin{titlepage}
\begin{flushright}
MAD-TH-15-07
\end{flushright}

\vfil

\begin{center}

{\bf \Large
Intersecting D3/D3' system at finite temperature
}

\vfil

William Cottrell, James Hanson, Akikazu Hashimoto,\\
 Andrew Loveridge, and Duncan Pettengill

\vfil

Department of Physics, University of Wisconsin, Madison, WI 53706, USA

\vfil

\end{center}

%%%%%%%%%%%%%%%%%%%%%%%%%%%%%%%%%%%%%%%%%%%%%%%%%%%%%%%%%%%%%%%%%%%%%%%%%%%%%%%%%%%%%%%
\begin{abstract}
\noindent We analyze the dynamics of intersecting D3/D3' brane system
overlapping in 1+1 dimensions, in a holographic treatment where $N$
D3-branes are manifested as anti-de-Sitter Schwartzschild geometry,
and the D3'-brane is treated as a probe. We extract the thermodynamic
equation of state from the set of embedding solutions, and analyze the
stability at the perturbative and the non-perturbative level. We
review a systematic procedure to resolve local instabilities and
multi-valuedness in the equations of state based on classic ideas of
convexity in microcanonical ensumble. We then identify a run-away
behavior which was not noticed previously for this system. 

\end{abstract}
%%%%%%%%%%%%%%%%%%%%%%%%%%%%%%%%%%%%%%%%%%%%%%%%%%%%%%%%%%%%%%%%%%%%%%%%%%%%%%%%%%%%%%%%%
\vspace{0.5in}

\end{titlepage}
\renewcommand{\baselinestretch}{1.05}  %Line spacing
%%%%%%%%%%%%%%%%%%%%%%%%%%%%%%%%%%%%%%%%%%%%%%%%%%%%%%%%%%%%%%%%%%%%%%%%%%%%%%%%%%%%%%%%%%%%%

\section{Introduction}

Recently in \cite{Mintun:2014aka}, a surprising subtelty was identified in a deceptively simple system of intersecting D-branes.  Consider a system consisting of a D3 and a D3' brane in type IIB string theory, oriented according to  
\be
\begin{tabular}{c||cccccccccc}
       & 0& 1  & 2& 3& 4& 5& 6& 7& 8& 9 \\
       \hline
D3 & $\circ$ &  $\circ$ & $\circ$ &  $\circ$ &    &   &    &   &   &     \\
D3' & $\circ$ &   && $\circ$ & $\circ$ & $\circ$ &    &   &    \\
\end{tabular} \label{orientation}
\ee
and separated by a finite distance in the $x_9$ direction.

Such a system preserves 8 supercharges. The low energy open string
degrees of freedom can easily be enumerated as consisting of
\begin{enumerate}
\item An ${\cal N}=4$ $d=4$ $U(1)$ gauge theory living on D3
\item An ${\cal N}=4$ $d=4$ $U(1)'$ gauge theory living on D3'
\item Two sets of hypermultiplets $B$
and $C$, arising from ${\cal N}=2$ $d=4$
hypermultiplets, dimensionally reduced to $d=1+1$ dimensions and
charged as a bi-fundamental under $U(1) \times U(1)'$
\end{enumerate} 
Explicit coupling between these states was worked out in
\cite{Constable:2002xt}.  This system, consisting entirely of D3
branes in type IIB string theory, is manifestly self-dual under
$S$-duality.

On the first pass, there appears to be no obstruction to taking the
zero slope limit $\alpha' \rightarrow 0$ as long as one scales the
distance separating the D3 branes to be of order
\be \Delta x_9 = \alpha' V \ee
for some $V$ with dimension of mass. This then should give
the mass of the $B$ and $C$ fields corresponding to the lowest energy
33' strings.

As was pointed out in \cite{Mintun:2014aka}, this system exhibits a
subtle paradox.  A state with a single quantum of the $B$ or $C$
fields should exist as a BPS state in the spectrum of the theory, so
its magnetic dual must also exist as a BPS state in order to be
consistent with $S$-duality. Such a state should arise as a soliton of
the field theory at hand. However, the soliton in question doesn't
appear to exist; something must therefore be wrong with the
assumptions being made about the system.

The resolution proposed by \cite{Mintun:2014aka} was that the zero
slope limit failed to achieve decoupling. They then argued that a
soliton does exist for a suitably modified effective field theory
which contains singularities, signaling the need to include additional
UV degrees of freedom. By considering full string theory as a UV
completion, for instance, the effective field theory can be
regularized, and the soliton can be constructed as the expected
magnetic dual state.

Following the work of \cite{Mintun:2014aka}, a simple generalization
in the brane construction was considered in \cite{Cottrell:2014ura}. This
construction involved also scaling the angle between the D3-branes as
$\theta \sim \alpha' c$, introducing a new scale $c$ with dimension of
mass squared. This gave rise to a tower of states of which $B$ and $C$
are the lightest \cite{Hashimoto:1997gm}. In that setup, the decoupled
theory in the zero slope limit is perfectly sensible, and supports the
magnetic monopole soliton. One therefore learns that one can complete
the effective field theory of \cite{Mintun:2014aka} more economically
than by invoking full string theory.

Taking the limit $c\rightarrow \infty$ while keeping $V$ fixed in the
construction of \cite{Cottrell:2014ura} essentially amounts to
recovering the naive zero slope limit considered in
\cite{Mintun:2014aka}. The techniques employed in
\cite{Cottrell:2014ura} were not particularly effective for studying
this limit, but in \cite{Cottrell:2014nea}, we introduced another
variation in the setup where we replaced the D3 with a stack of $N$
D3's, so that we have as a gauge group $U(N) \times U(1)'$. This
allows us to analyze the system in a strong coupling limit where the
$N$ D3-branes are replaced by their $AdS_5 \times S_5$ dual, and the
D3' is treated as a probe. We then considered the magnetic soliton
realized as a bion \cite{Callan:1997kz} melting into the horizon along
the lines of \cite{Schwarz:2014rxa}. It is then possible to see how in
the $c \rightarrow \infty$ limit, the magnetically charged bionic
soliton delocalizes and decouples as a normalizable state in the $c
\rightarrow \infty$ limit.

The behavior of the D3/D3' intersection is so counterintuitive that we
probably have not yet seen the last word regarding this system. One
direction which seems potentially interesting is to explore the
thermodynamic behavior of this model. An elegant way to approach this
issue incorprating the effects of interactions is to use gauge gravity
correspondince, treating the D3'-brane as a probe, along the lines of
\cite{Cottrell:2014nea}. Working in finite temperature then amounts to
studying the embedding of a probe D3'-brane in an AdS-Schwartzschild
background.

Problems of this type where a D$p'$-brane probe is embedded into
finite temperature D$p$-brane geometries have been considered
extensively. Most of these works were in the context of exploring
meson dynamics in holographic QCD
\cite{Kruczenski:2003be,Kruczenski:2003uq,Mateos:2006nu,Mateos:2007vn,Myers:2006qr,Benincasa:2009be,Myers:2007we,Hoyos:2006gb}. It
was observed that these brane embeddings undergo a phase transition
in which they penetrate the black hole horizon. This phase transtion was
interpreted as ``melting'' of mesons, which was supported by
subsequent analysis of the spectrum of fluctuations on the probe
brane. The behavior near criticality for the melting transition also
has a rich structure which was noticed even earlier in the context of
domain walls \cite{Christensen:1998hg,Frolov:1998td}. This analysis
was carried out in various combinations of D$p$-brane background and
D$p'$-brane probes, and the general behavior of the system is
dimension independent at least in a broad brush perspective.

In these class of problems, one generally enumerates the classical
solutions to the equation of motion corresponding to static brane
embedding configurations. The embeddings are characterized by a control
parameter corresponding to quark mass, and have a definite order
parameter corresponding to the quark condensate. The static solutions
constrain the equation of state relating the quark condensate to
quark mass. Treating these parameters as thermodynamic
quantities, one can explore issues such as thermodynamic stability and
hydrodynamic limits. Indeed, these system generally exhibits
instabilities and multi-valued equation of states as were observed,
for instance, in
\cite{Cvetic:1999rb,Cvetic:1999ne,Chamblin:1999hg,Chamblin:1999tk}.

The case of D3'-brane probe in the background of D3 branes oriented according to (\ref{orientation}), however, is somewhat special. This is the case that was singled out in Reference [21] of \cite{Mateos:2006nu}. There are two concrete senses in which the D3/D3' system stands out. The brane embedding is characterized by a scalar on the D3'-brane world volume which happens to saturate the Breitenlohner-Freedman $m^2 \ge -1$. \cite{Constable:2002xt}. As a result, the operators corresponding to these fields have scaling dimension
\be \Delta = {d \mp \sqrt{d^2 + 4 m^2} \over 2} = 1 \ee
which is degenerate for $d=2$ and $m^2=-1$. This gives rise to
logarithmic factors in the scaling of the embedding solution near the
boundary, and exchange the roles of control and order parameters in
the holographic dictionary as we will ellaborate further below. Also,
the holographic renormalization procedure requires introducing an
anomalous scale which can affect certain physical observables
\cite{Karch:2005ms}.

The thermodynamics of D3/D3' system have been analyzed previously
\cite{Hung:2009qk} although in that threatment, the control parameter
of the embedding was treated as being fixed in defining the ensumble. The
analysis of \cite{Hung:2009qk} was primarily focused on establishing
the existence of a robust zero mode in the longitudional electric
fluctuations and its implication for charge transport behaviors.

In this article, we re-examine the thermodynamics of D3/D3' system
with emphasis on understanding the thermodynamic stability issue of
the quark condensate order parameter. We will explore the stability
both at the perturbative and the non-perturbative level, and argue
that the phase diagram of the system looks somewhat different than
what was suggested in \cite{Hung:2009qk}. For now, we will focus
primarily on zero charge embeddings. The extention of this analysis
including charges and chemical potential will be reported in a
separate publication.

\section{D3'-brane probing finite temperature anti de Sitter space}

In this section, we will review the analysis of embedding D3'-branes
in the near horizon geometry of $N$ D3-branes at finite temperature.
Similar analysis can be found extensively in the literature, but it is
useful to formulate it here to make the notation and
conventions explicit.

\begin{figure}
\centerline{\includegraphics[width=3in]{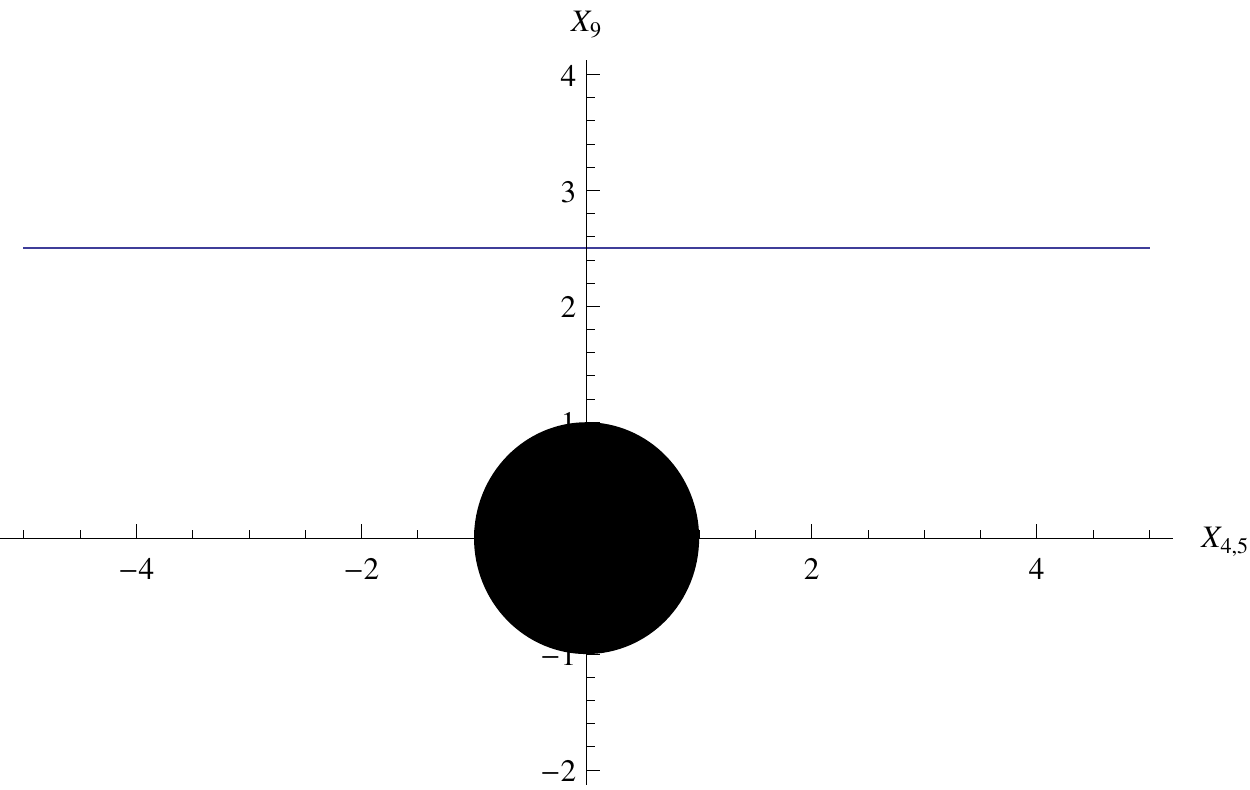}}
\caption{ A schematic illustration of a flat brane embedding. This embedding is static in the absence of the black hole. We are interested in how this embedding is deformed when the black hole is introduced. 
\label{figa}}
\end{figure}

We begin by writing the supergravity solution corresponding to a stack of $N$ D3-branes at finite temperature in type IIB supergravity \cite{Horowitz:1991cd}
\be ds^2 = -H^{-1/2} ( - f dt^2 + d \vec x^2) + H^{1/2} (f^{-1} dr^2 + r^2 d \Omega_5^2 ) \ee
with 
\be H = 1 + {R^4 \over r^4}, \qquad R^4 = 4 \pi g_s N \alpha'^2 = \lambda \alpha'^2 \ , \ee
and
\be f = 1 - {r_s^4 \over r^4} \ . \ee
By standard arguments, the temperature $T$ and the horizon radius $r_s$ are related by 
\be T = {1\over \pi r_s \sqrt{H(r_s)}} = {U_s \over  \pi \sqrt{\lambda}}.  \label{TU} \ee

We will take the decoupling limit by sending $\alpha' \rightarrow 0$
keeping $U \equiv r /\alpha'$  and $T$ fixed. This will also keep $U_s = r_s
/\alpha'$ fixed and finite.

At zero temperature, brane configuration oriented as (\ref{orientation}) and illustrated in figure \ref{figa} is a consistent static solution. We are interested in how such an embedding is deformed when $T$ is no longer zero. 

We will parameterize the six dimensions transverse to the $N$ D3-branes  as
\beq x_4 &=& r \sin \theta_1 \sin \theta_2 \sin \theta_3 \sin \theta_4 \cos \theta_5\\
x_5 &=& r \sin \theta_1 \sin \theta_2 \sin \theta_3 \sin \theta_4 \sin \theta_5\\
x_6 &=& r \sin \theta_1 \sin \theta_2 \sin \theta_3 \cos \theta_4 \\
x_7 &=& r \sin \theta_1 \sin \theta_2 \cos \theta_3\\
x_8 &=& r \sin \theta_1 \cos \theta_2 \\
x_9 &=& r \cos \theta_1  \ . 
\eeq
We can then treat 
\be t, \qquad z = x_3, \qquad r, \qquad \phi =\theta_5 \ee
as the world volume coordinates in static gauge for the embedding, and denote
\be d^4 \sigma = dt\, dz \, dr\,  d \phi\ r \ee
and treat $\theta_1$, $\theta_2$, $\theta_3$, and $\theta_4$ as
parameterizing the embedding. We will restrict our attention to
configurations where the gauge field on the world volume of D3' is
trivial at first. The system is invariant under $SO(4)$ rotational
invariance acting on $(x_6,x_7,x_8,x_9)$, and it turns out to be
dynamically consistent to set all but one of the four coordinates to
zero. This is equivalent to setting
\be\theta_2=\theta_3 = \theta_4 = {\pi \over 2} \ee
and treating $\theta=\pi/2-\theta_1$ as the only relevant field
variable.  Restricting to embeddings which are invariant under
translation in $t$ and $z$ directions, the resulting effective
action is
\be S  = {1 \over (2 \pi)^3 \alpha'^2 g_s} \int d^4 \sigma\    \cos (\theta(r)) \sqrt{1+  r^2 f(r) \theta'(r)^2} \ . \label{effective}\ee
Note in particular that the dependence on warp factor $H$ dropped out
completely from the action.\footnote{The warp factor is nonetheless
  relevant for the formula relating temperature to horizon radius
  (\ref{TU}). The warp factor also enters in the computation of quasinormal modes in appendix \ref{appB}.}

In the zero slope limit,  one sees that $\alpha'$ scales out of the action
\be S_{DBI}  = {1 \over (2 \pi)^3 g_s} \int d^4 \Sigma\,    \cos (\theta(U)) \sqrt{1+ \left(U^2 -{U_s^4 \over U^2}\right)  \theta'(U)^2}  \label{nonlin}  \ee
with 
\be d^4 \Sigma = dt\, dz \, dU \, d \phi\, U \ . \ee

For the purpose of finding the embeddings which extremizes this action, it is convenient to scale $U_s$ out by defining
\be u = {U \over U_s} \ee
so that the action takes the form
\be S  = {U_{s}^{2} \over (2 \pi)^2 g_s} \int d^2x \,  du \, u\,     \cos (\theta(u)) \sqrt{1+ \left(u^2 -{1 \over u^2}\right)  \theta'(u)^2}  \label{nonlin2} \,  \ee
where $L$ is the volume of the $x_1$ coordinate and $T$.

The task at hand now is to solve the equation of motion obtained by
varying (\ref{nonlin2}) which is a non-linear second order
differential equation for $\theta(u)$ for $u$ taking values in the
range $1 \le u \le \infty$. For large values of $u$, $\theta(u)$ approaches  zero and one can show that the solution can be parameterized in the form
\be \theta(u) = \left( {c \log(u) \over u} + {m \over u}\right) \label{cmu} \ee
where we denote dimensionless integration constants\footnote{This $c$
  is unrelated to the $c$ parameterizing the tilt of D3 relative to
  D3' in the notation of \cite{Cottrell:2014nea}.} $c$ and $m$
following the convention of \cite{Hung:2009qk}. Near $u=1$, we impose
the regularity condition which constrains $c$ as a function of $m$.

\begin{figure}
\centerline{\includegraphics[width=4in]{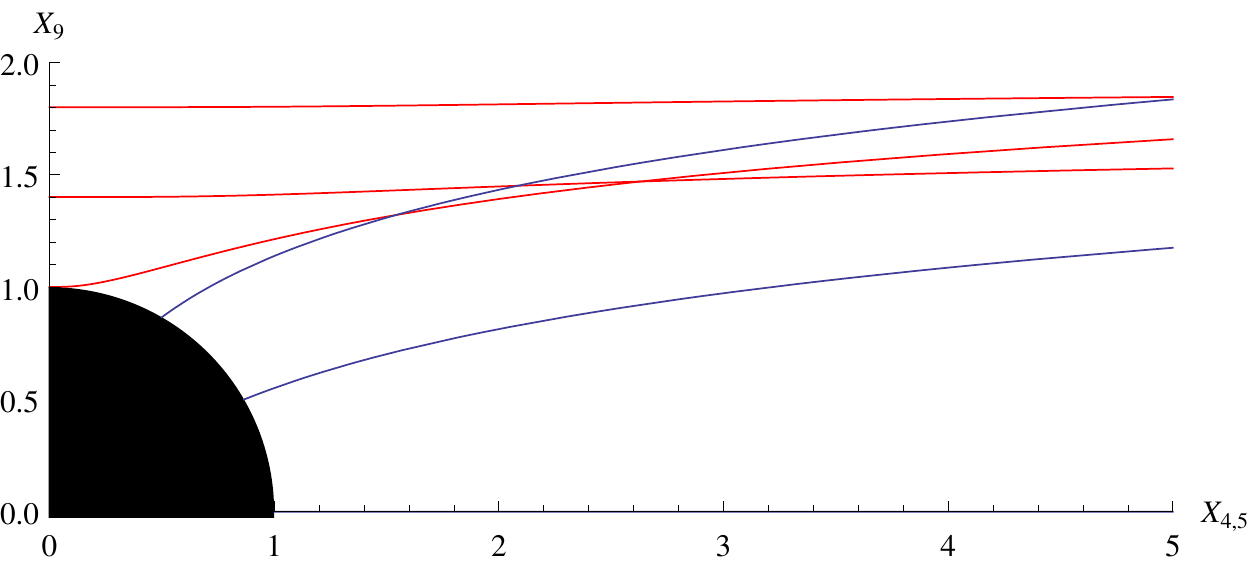}}
\caption{Black hole embedding illustrated in blue and Minkowski embeddedings illustrated in red.
\label{figd}}
\centerline{\includegraphics{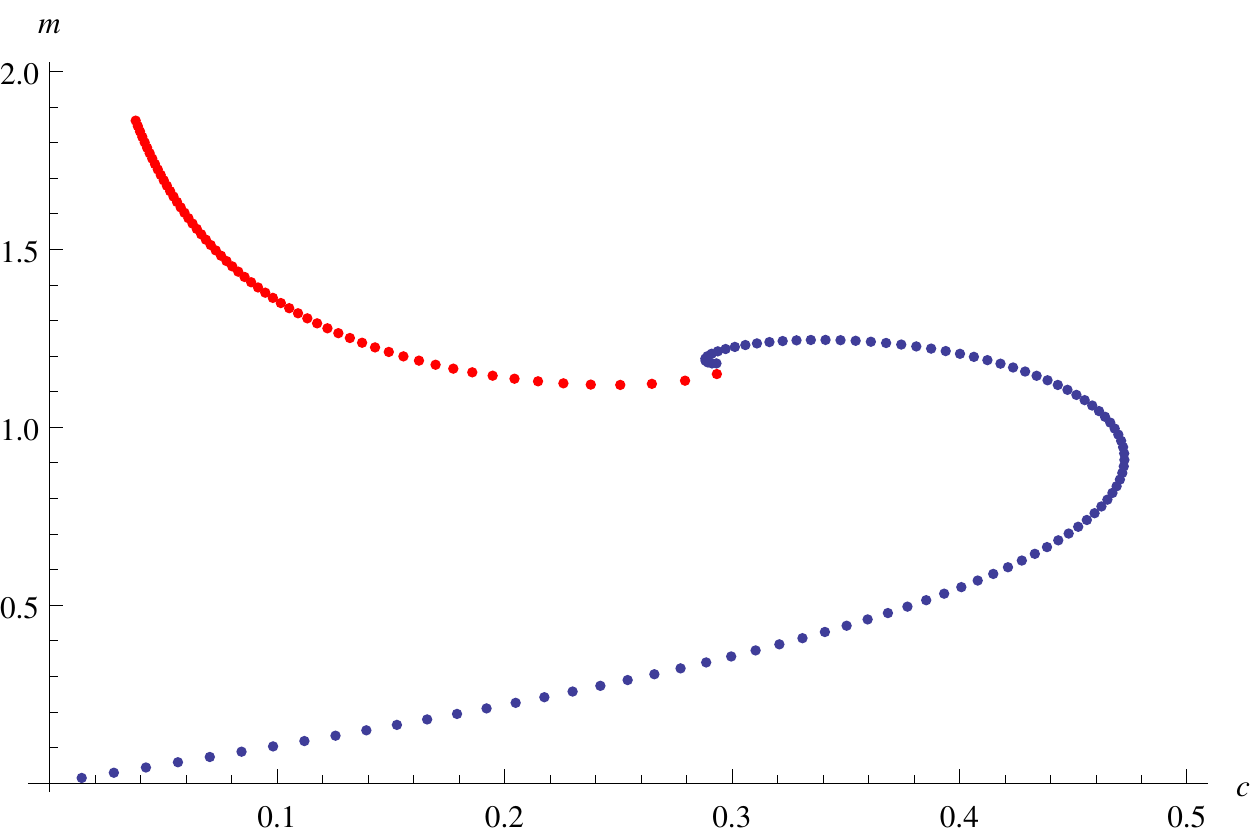}}
\caption{$(c,m)$ for Minkowski (red) and black hole (blue) embeddings of D3'-probe. Each dot corresponds to numerical solutions we found with initial conditions for $\theta(U)$ specified either at the horizon $U=U_s$ for the black hole embeddings, or by fixing $\theta'(U)$ at $\theta(U)=\pi/2$  for the Minkowski embeddings.
\label{figf}}
\end{figure}

The actual solutions satisfying these boundary conditions have to be
obtained numerically. The general feature of the solutions we obtained
is illustrated in figure \ref{figd}. There are two different class of
solutions depending on whether the brane penetrates the black hole
horizon or not. The ones which do not, known as Minkowski
embeddings, are illustrated in red. The ones which do, known as
black hole embeddings, are illustrated in blue.  Each of these
solutions have a definite value of $c$ and $m$. The set of $(c,m)$
computed for these solutions are illustrated in figure \ref{figf}.

There are a number of features that are worth noting in figures
\ref{figd} and \ref{figf}. First, $m$ is not single valued as a
function of $c$.  Closely related is the fact that there is a maximum
value of $c$ where $dc/dm=0$. Also, note that the embedding exhibits a
self similar critical structure when the Mikowski and black hole
embeddings meet. This was a feature originally noted
\cite{Christensen:1998hg,Frolov:1998td} and further ellaborated in the
context of D$p$/D$q$ system in \cite{Mateos:2006nu,Mateos:2007vn}. It
signals that there is a first order ``meson melting'' phase transition near the
self-similar critical point. In this article, we will have more to say
about the critical behavior at $dc/dm=0$ than at the self similar
point.

\section{Thermodynamics and Holography of D3/D3' system}

In this section, we elaborate on the thermodynamic and holographic
interpretations of embedding solutions illustrated in figures
\ref{figd} and \ref{figf}.

First, it should be noted that all embeddings illustrated in figures
\ref{figd} and \ref{figf} are asymptotically $AdS_3 \times S_1$. The
world volume degrees of freedom on D3'-brane will have a holographic
interpretation as a field theory in 1+1 dimensions.

The embedding field $\theta(U)$ will be associated to an operator of dimension $\Delta = 1$ to be associated with quark bilinar $\bar \psi \psi$ via the standard holographic dicationary which needs to be stated with some care because $\Delta$ is degenerate. Let us ellaborate on this matter further.

To interpret the system holographically, it is awkard to scale out $U_s$ since the holograhpic dictionary should be independent of $U_s$. Let us therefore parameterize the asymptotic behavior of the $\theta(U)$ in the form
\be \theta(U) = {C \log(U/U_*) \over U} + {M \over U} \label{CMU} \ee
where once again adapting the notation of \cite{Hung:2009qk}, $C$ and
$M$ have dimension of mass, whereas $U_*$ is an arbitrary scale with
dimension of mass which one must introduce in order to make sense of the
argument of the logarithm. An astute reader should notice at this
stage that there is some ambiguity in how $M$ is defined since
changing $U_*$ has the effect of shifting $M$. It would therefore be
essential to understand if and how $U_*$ affects physical observables
(or not.) Several related issues will arise in the discussion below
and we will be doing due diligence to track these issues.

We begin by recalling the standard formulation of the holograhpic dictionary (see e.g.\ \cite{D'Hoker:2002aw} for a review) that
\be Z_{bulk}[C(x)] = \langle e^{\int d^d x \, C(x) {\cal O}(x)} \rangle_{boundary} \label{AdSdict0} \ee
where the left hand side of the equality describes a path integral for bulk fields such as $\theta(U)$ carried out in such a way that $\theta(U)$ asymptotes to
\be \theta(U) \sim {C(x) \log(U/U_*) \over U} \ee
as the boundary is approached by taking $U \rightarrow \infty$. This path integral and boundary condition as formulated is independent of $U_*$.  One might formulate the right hand side for the $\theta(U)$ field to take the form
\be Z_{bulk}[C(x)] =  \int [D \theta(U,x)]_{C(x)} e^{-S_{DBI}[\theta(U,x)]} \ee
where $S_{DBI}$ is the action given in (\ref{nonlin}), and the
boundary condition for $\theta(U)$ is referenced implicitly in the
specification of the measure. One can apply the saddle point
approximation to identify the domimant contribution to this path
integral, which simply amounts to evaluating the action for the
solution to the equations of motion enumerated in figures \ref{figd}
and \ref{figf}. As is typical in these computation, however, the
action formally diverges, and a renormalization is required to define
the bulk side of the correspondence unambiguously. For the D3/D3'
system, this was worked out explicitly in (6.2)--(6.4) of
\cite{Karch:2005ms}. We are simply instructed to add the holographic
renormalization counter-term
\be S_{CT} = \left.{1  \over (2 \pi)^2 g_s}\int d^2 x\,     U^2 \left(-{1 \over 2}   + {1 \over 2} \left(1 - {1 \over \log(U/U_{CT})}\right)  \theta(U)^2 \right) \right|_{U=U_{UV}}\ . 
\ee
With this counter-term included,
\be Z_{bulk}[C(x)] =  \int [D \theta(U,x)]_{C(x)} e^{-(S_{DBI}[\theta(U,x)]+S_{CT}[\theta(U,x)])} \label{AdSdict} \ee
where $U_{UV}$ is the ultra-violet cut-off scale, whereas $U_{CT}$ is a new scale that is required in order to make the logarithm appearing in the counter-term make sense. This expression is finite in the $U_{UV} \rightarrow \infty$ limit. However, the dependence on  $U_{CT}$ which characterizes the renormalization scheme survives and should a priori be treated as independent of $U_*$ and $U_s$. 
A useful feature to isolate in the holographic dictionary is the prescription to extract the expectation value of the operator dual to $\theta(U)$. This is to be derived by varying the logarithm of $Z[C(x)]$ with respect to $C(x)$. By manipulating $S_{DBI}+S_{CT}$, one finds that
\be \langle {\cal O}(x) \rangle = - {1 \over (2 \pi)^2 g_s} ( M(x) - C(x) \log(U_*/U_{CT}) ) \ . \ee

It is also useful to infer the values of  $(C,M)$ for the solutions enumerated in figures \ref{figd} and \ref{figf} which are parameterized in terms of $(c,m)$. By simply relating  (\ref{cmu}) to (\ref{CMU}), we find that
\be C = U_s c, \qquad M = U_s \left(m - c \log(U_s/U_*) \right) \ . \ee

In terms of $(c,m)$, we can write
\be \langle {\cal O}(x) \rangle = -  {1 \over (2 \pi)^2 g_s} U_s (m - c \log(U_s/U_{CT})) \ee
whose significance is the fact that the dependence on $U_*$ has dropped out. However, the dependence on $U_s$ and $U_{CT}$ remains.

At this point, we can also compute the free energy by evaluating the action with time compactified on a circle of radius $1/2\pi T$
\be G(C,T) = - T \log(Z[C]) \ee
or by computing
\beq G(C,T) &=& -{L \over (2 \pi)^2 g_s} \int_0^C dC' (M(C) - C' \log(U_*/U_{CT})) \cr
&=&
-{L U_s^2 \over (2 \pi)^2 g_s} \int_0^c dc' (m(c) - c' \log(U_s/U_{CT})) \eeq
which gives an equivalent $U_*$ independent result. Since $U_*$ is essentially unphysical, it is conveninet to set $U_*=U_{CT}$ so that
\be \langle {\cal O}(x) \rangle = {1 \over (2 \pi)^3 g_s} M   \ee
for the remainder of this paper.

\begin{figure}
  \centerline{\includegraphics[width=3in]{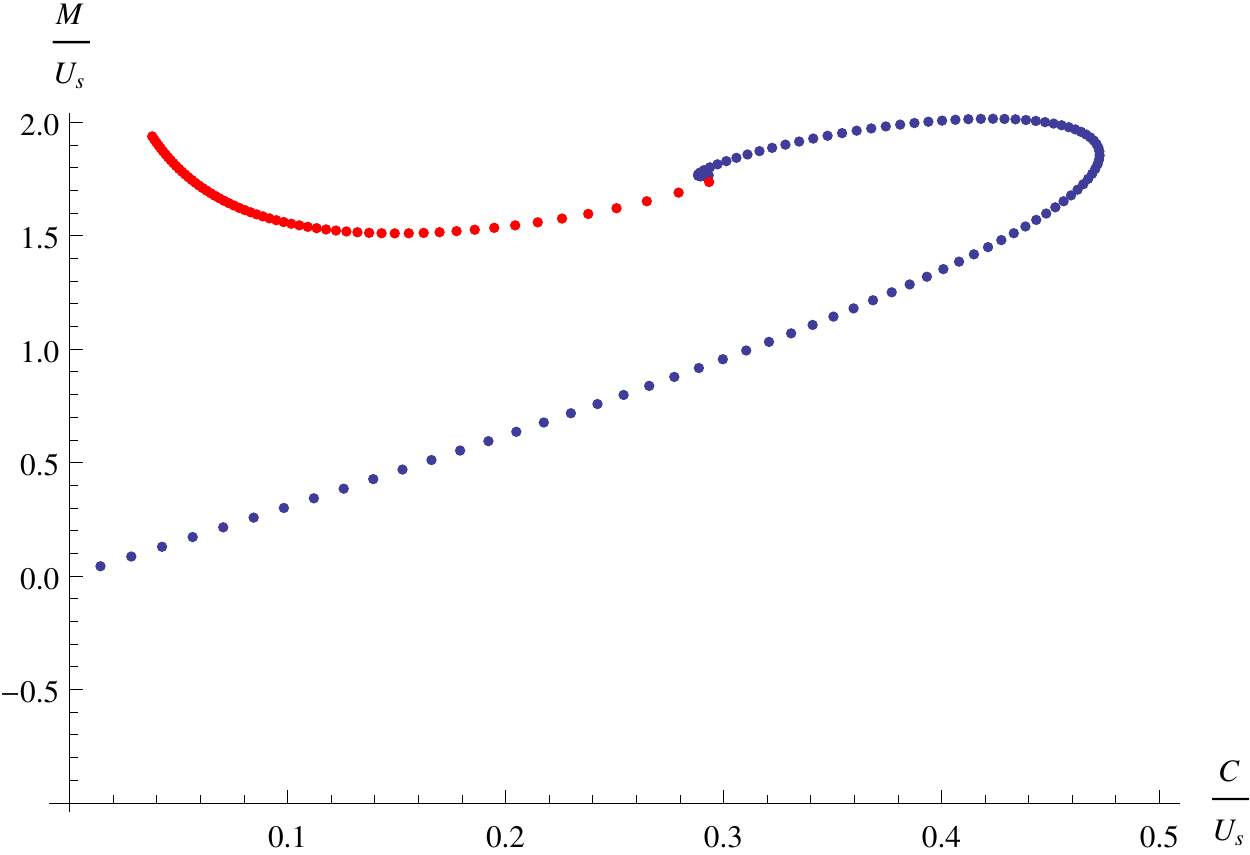} \includegraphics[width=3in]{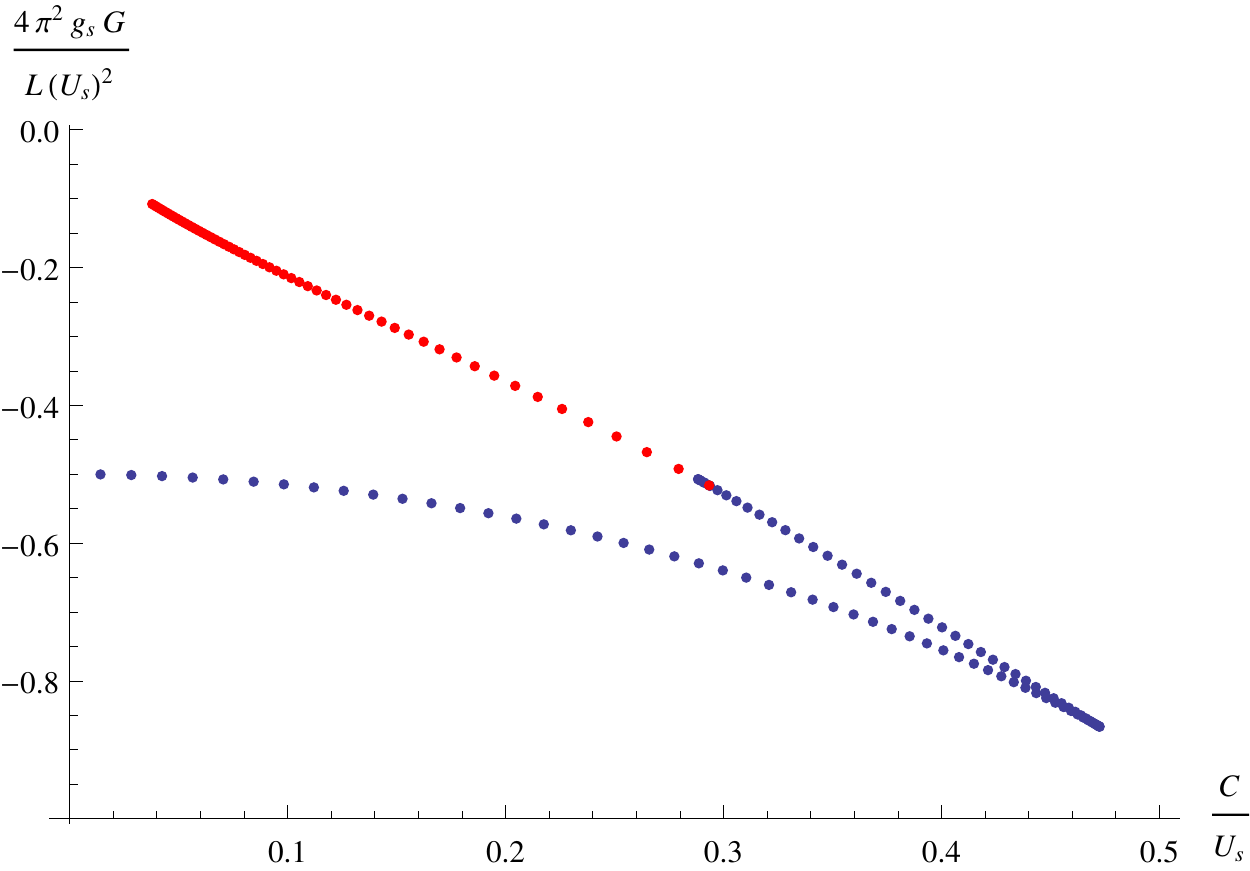}}
  \centerline{\includegraphics[width=3in]{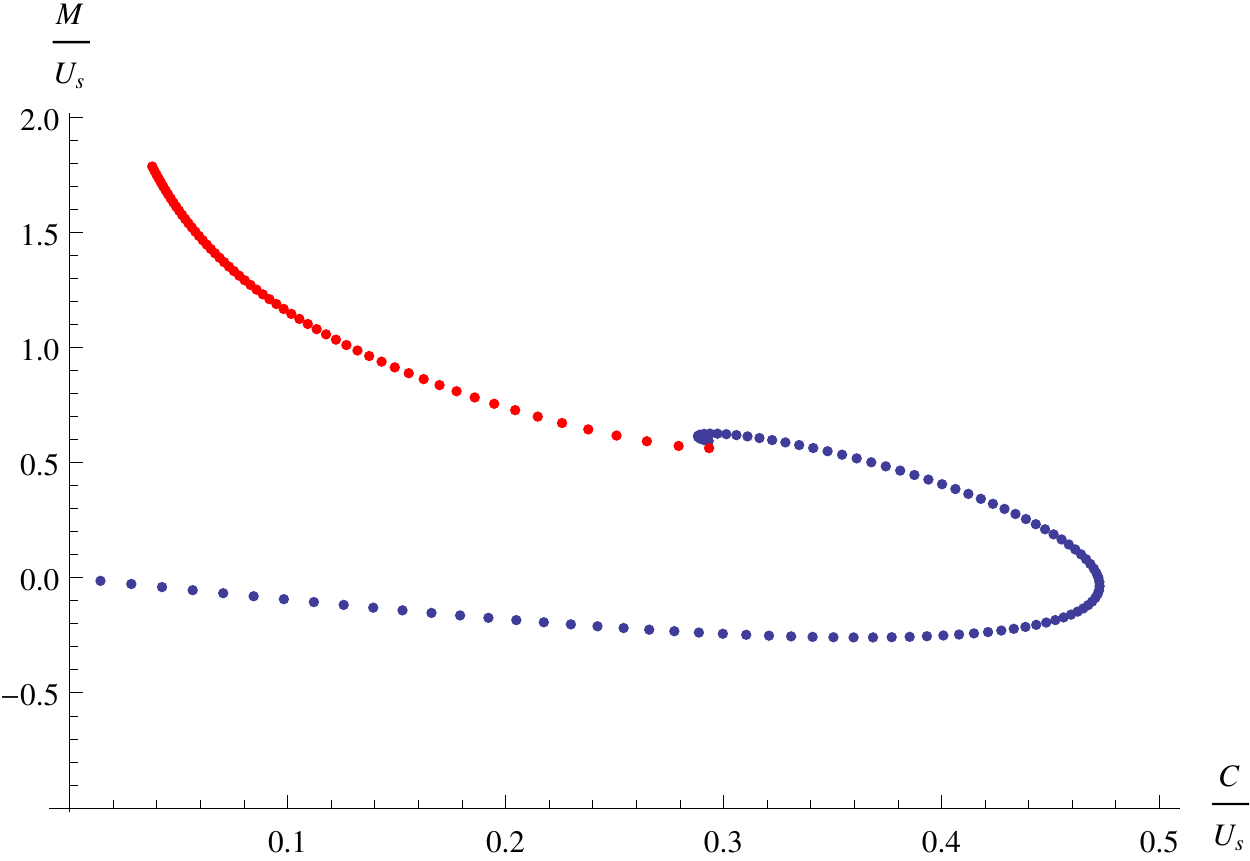} \includegraphics[width=3in]{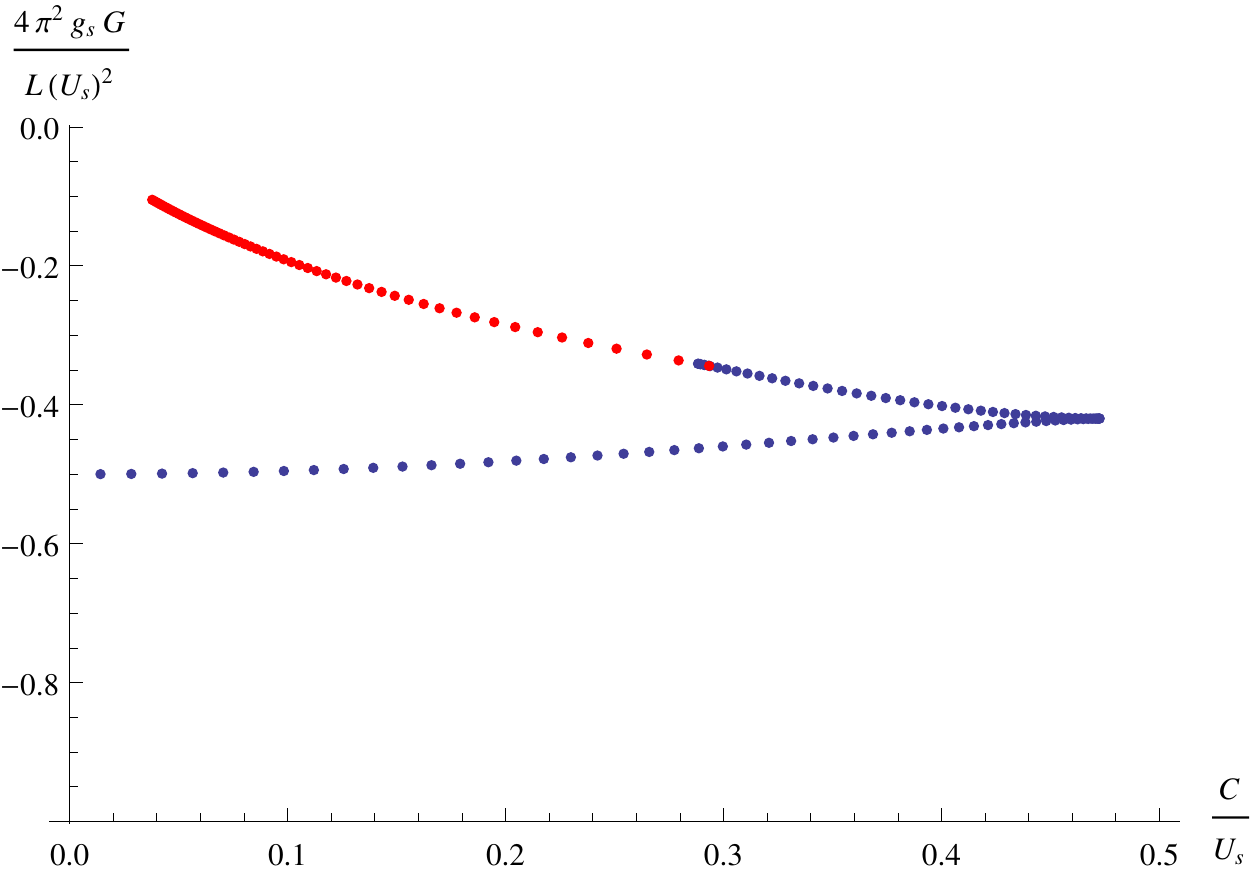}}
\caption{$M(C,T)$ and $G(C,T)$ as a function of $C$ for $\log(U_s/U_{CT}) = -2$ (top) and $\log(U_s/U_{CT}) = 2$ (bottom) \label{figq}}
\end{figure}

We are also now in the position to display thermodynamic data such as the equation of state $M(C,T)$ and the free energy $G(C,T)$ for various fixed values of $T/U_{CT}$. Few examples are illustrated in figures \ref{figq}.

\begin{figure}
  \centerline{\includegraphics[width=4in]{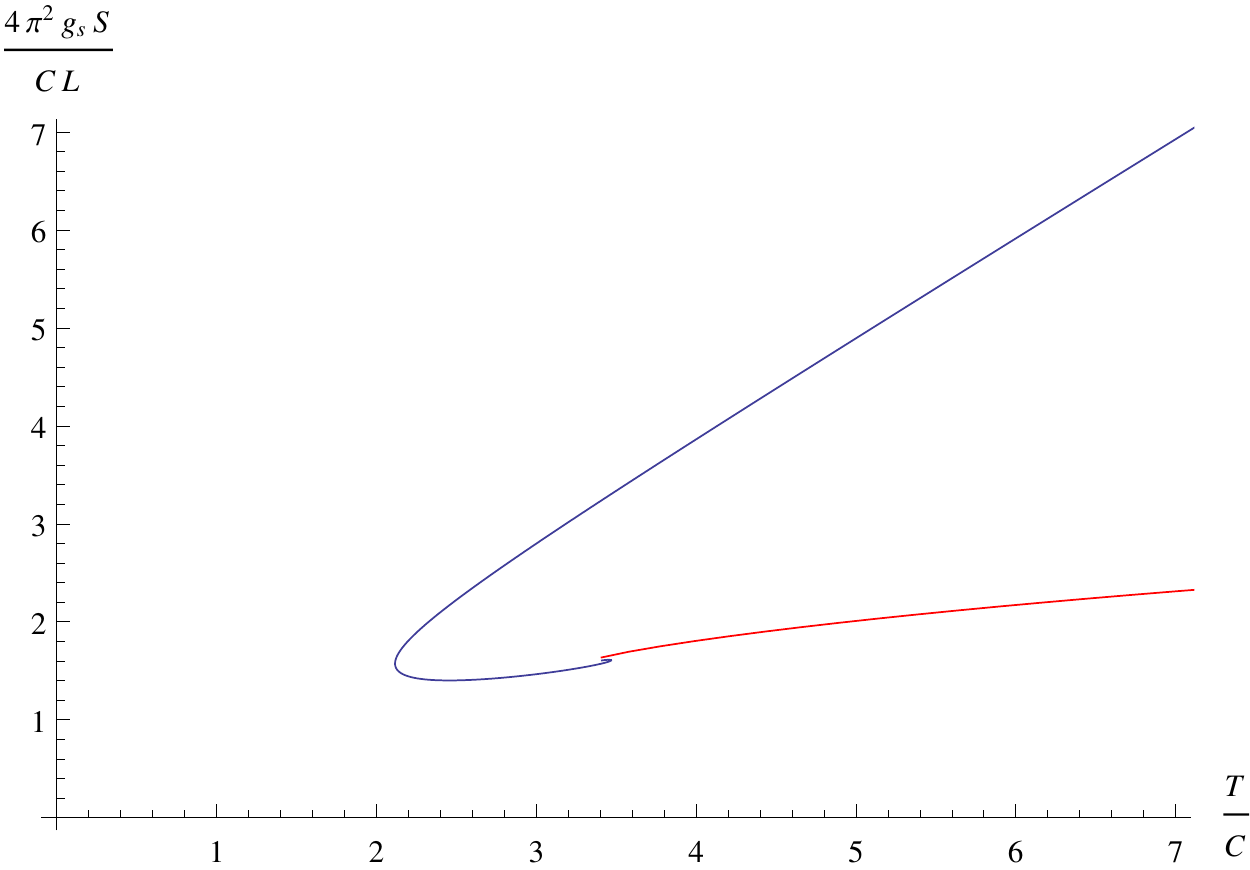}}
  \caption{$S(C,T)$ as a function of $T$ for fixed $C$. This curve turns out to not depend on $U_{CT}$. \label{figr}}
\end{figure}

It is also straight forward to infer quantity such as
\be S(C,T) = -\left.{\partial G \over \partial T}\right|_{C} = {L U_s \over (2 \pi)^2 g_s}  \left(2 \int_0^{c} m(c) dc' - cm - {1 \over 2} c^2 \right) \ee
which happens to be independent of $U_{CT}$ and plot it as a function of $T$ with $C$ fixed. We illustrate that in figure \ref{figr}. This is essentially equivalent to what is illustrated in figure 3.d of \cite{Hung:2009qk} except that we do not find any evidence of the dotted part of their graph. This view is supported also from the structure of equation of state illustrated in figure  \ref{figf} which has exactly two, not three, branches as $c$ approaches zero. We will ellaborate further on this point below.

\section{Physical interpretation of D3/D3' thermodynamics} 

In the previous section, we outlined the general features of the
D3/D3' system which can be presented in the thermodynamic context. The
control and order parameters, $(C,M)$ play a role very similar to
mechanical parameters $(P,V)$, magnetic parameters $(H,M)$, etc., up
to a sign which arises from conventions which are set for historical reasons.

There are two notable features about the equation of state and the
subsequent thermodynamics summarized in the previous section. One is
the fact that the scale $U_{CT}$ affects the equation of state which
is physically observable. The other is the fact that the equation of
state exhibits multi-valued-ness and regions of instability. This
latter issue is somewhat familiar from previous consideration of black
hole thermodynamics
\cite{Cvetic:1999rb,Cvetic:1999ne,Chamblin:1999hg,Chamblin:1999tk}. It
is generally stated that the van der Waals model of liquid-gas phase
transition is a prototype for understanding these issue. Nonetheless,
in the context of van der Waals model, it was the order parameter as a
function of the control parameter, $V(P)$ that was multi-valued,
whereas in the case of D3/D3' system, it is exactly the
opposite.\footnote{For an illuminating discussion of this very issue,
  see \cite{fisher}.} Another apparent paradox stems from the fact
that the sucsceptiblity
\be \chi_{\theta} = {d C \over dM} \ee
characterizing this system is explicitly dependent on $U_{CT}$ whereas dymamical features such as the poles of quasi-normal modes at fixed control parameter $C$ are manifestly independent of $U_{CT}$. The goal of this section is to clarify these issues.

Let us begin by recalling the classical thermodynamic perspective on stability. A useful quantity to consider is the effective potential of the order parameter obtained by Legendre transforming the free energy
\be F(M) = \left. G(C) + {L \over (2 \pi)^2 g_s } C M\right|_{G'(C)=-M} \ . \ee
One can also compute $F(M)$ in terms of the equation of state
\be F(M)= {L \over (2 \pi)^2 g_s} \int^M dM' C(M') \ee

\begin{figure}
  \centerline{\includegraphics[width=2.1in]{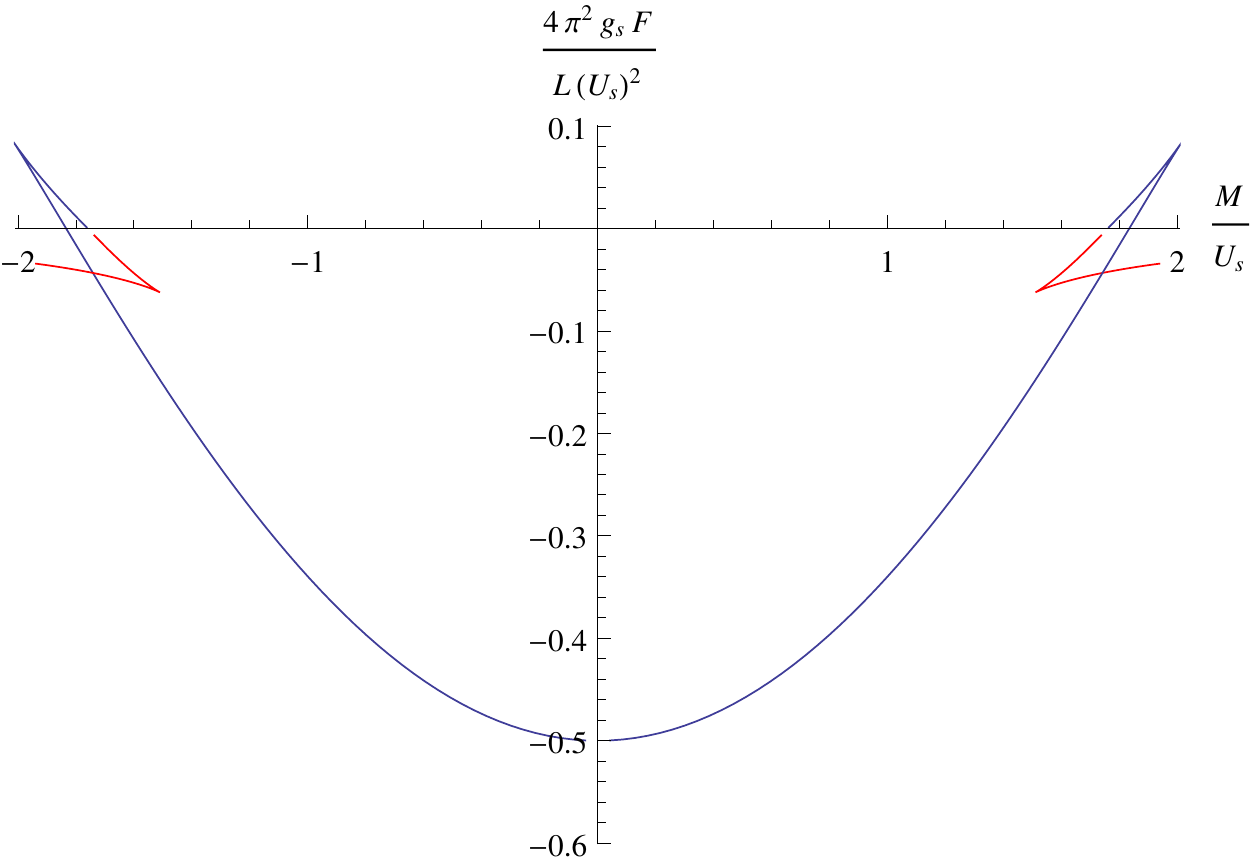}
  \includegraphics[width=2.1in]{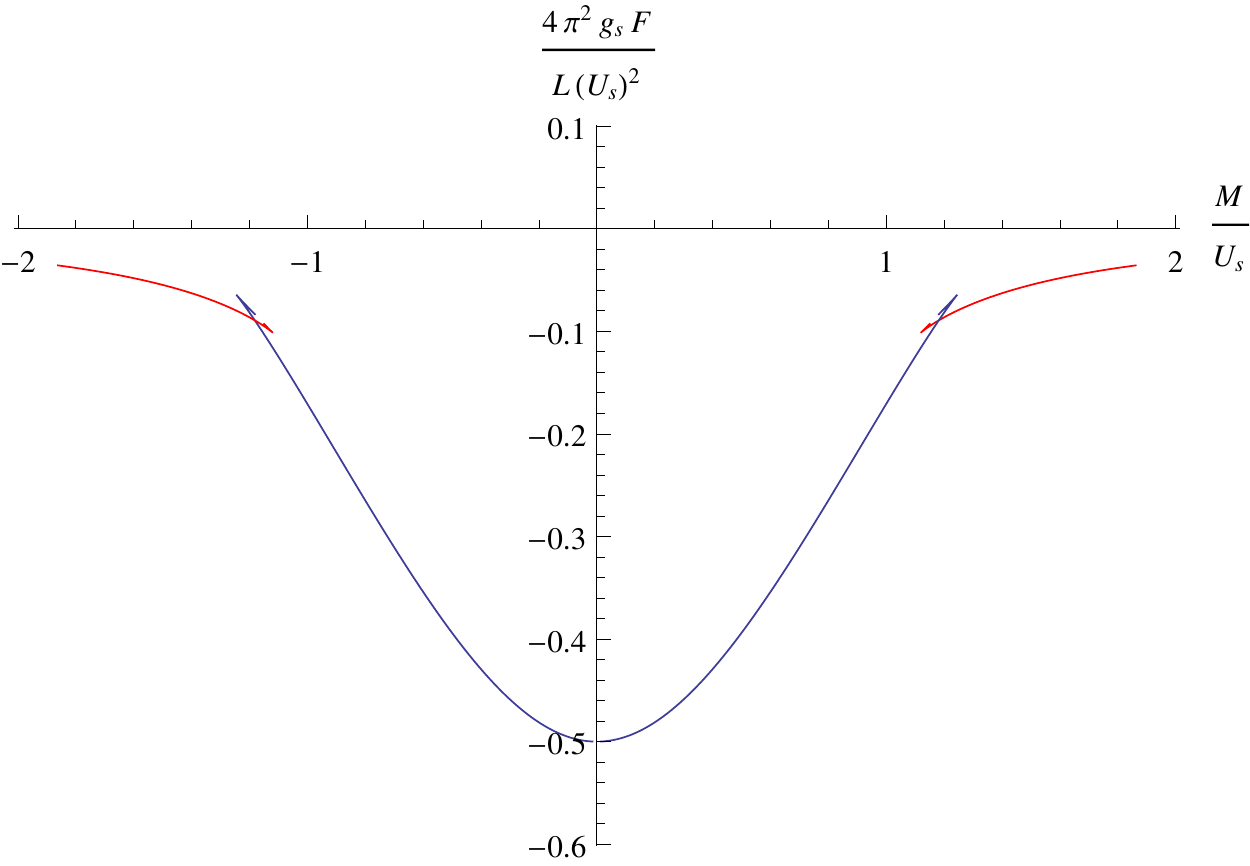}
  \includegraphics[width=2.1in]{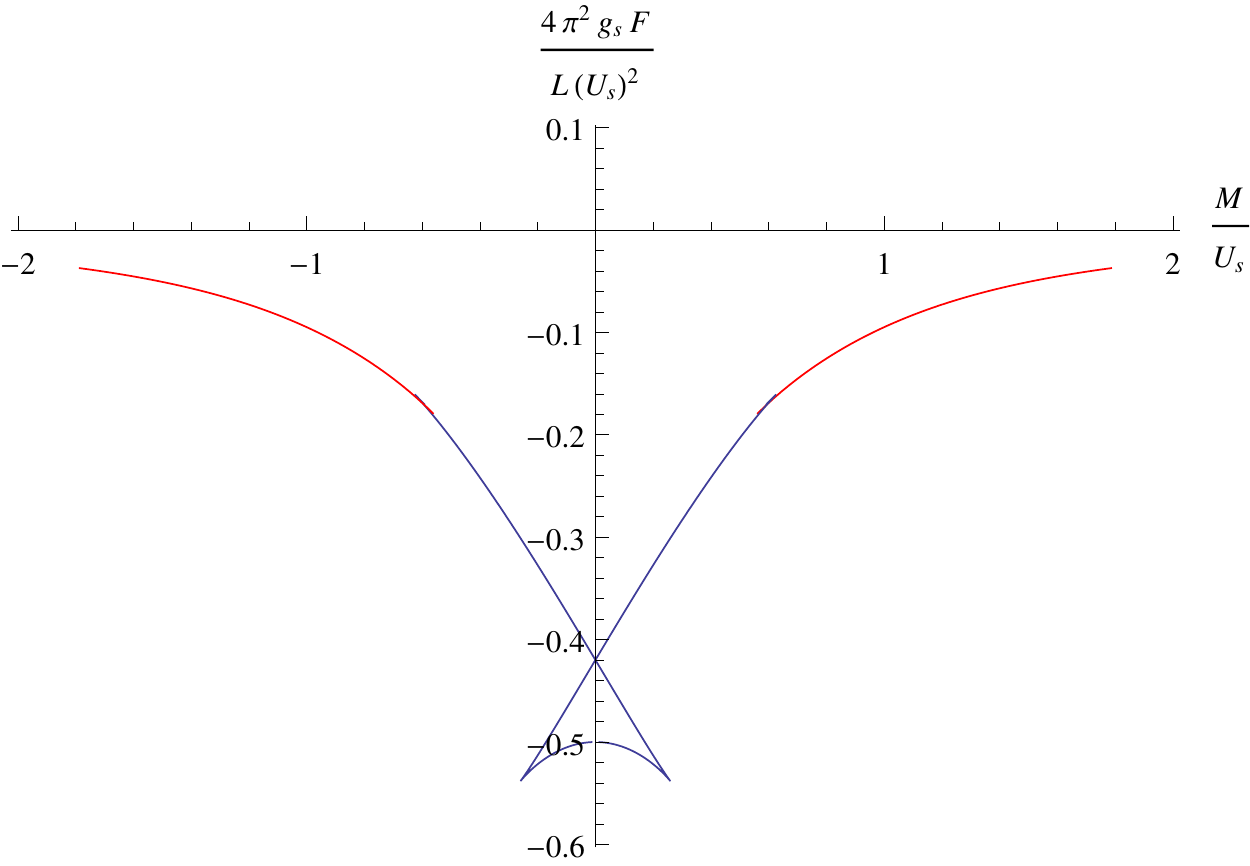}}
  \caption{$F(M)$ for $\log(U_s/U_{CT})$ taking values $-2$ (left), $0$ (center), and $2$ (right). For all of these cases, $F(M)$ asymptotes to a constant value 0 as $M$ is taken to  infinity.  \label{figs}}
  \end{figure}

Plotted as a function of $M$ with fixed values of $\log(U_s/U_{CT})$, they take the form illustrated in figure  \ref{figs}. For all of these figures, $M \rightarrow \infty$ corresponds to the Minkowski branch, and $F(M)$ approaches a constant, reflecting the fact that the area under the curve $C(M)$ is finite. 

The free energy $F(M)$ is an important physical quantity
characterizing the effective thermodynamic behavior of the order
parameter $M$. If a system with free energy $F(M)$ is brought to
contact with a reservoir with which the $M$-ness is freely exchanged,
the system will achieve equilibrium when $M$ minimizes the potential
\be F_{eff}(M) = F(M) - {L \over (2 \pi)^2 g_s} C_{ext} M \ee
where $C_{ext}$ is the control parameter conjugate to $M$ of the reservoir. One of course recognizes
\be G(C_{ext}) = \left. F(M) - {L \over (2 \pi)^2 g_s} C_{ext} M\right|_{F'(M) - C_{ext}=0} \ee
as the conjugate Gibbs Free energy when the $M$ that minimizes $F_{eff}(M)$ is substituted into $F_{eff}(M)$. From that point of view, it is natural to associate the susceptibility
\be \chi_\theta = {(2 \pi)^2 g_s  \over L } F''(M) = {d C \over dM} \ee
as parameterizing the stability of the system. The susceptibility $\chi_\theta$ is dependent on $U_{CT}$.

Let us now take a closer look at the form of $F(M)$ illustrated in figure \ref{figs} and make several observations. 
\begin{enumerate}
\item $F(M)$ is multi-valued over some range of $M$.
\item The susceptibility $\chi_\theta$ at $M=0$ (as well as other values of $M$) changes as $U_s/U_{CT}$ is varied, and can get negative, signalling an instability, for instance for large positve values of $\log(U_s/U_{CT})$.
\item The effective action $F_{eff}(M)$  is not concave everywhere and does not have global stable minima when $C_{ext}$ is non-vanishing. 
\end{enumerate}
Let us address each of these observations more carefully.
\subsection{Multi-valuedness of $F(M)$}

This issue is not too serious. The fact that there are multiple branches for some fixed value of $M$ (and $T$) simply reflects the fact that there are multiple thermodynamic states corresponding to these order and fixed parameters. However, in thermodynamics, one focuses on the dominant state in the ensumble, which is the one with the lowest free energy. So in reading figure \ref{figs}, one should simply trace the branch with smallest $F(M)$ for any fixed value of $M$, regardless of the discontinuities that might result. 

\subsection{Susceptibility and its dependence on $U_{CT}$}

This is an extremely important yet subtle issue. Taken at face value, it implies that the susceptibility and therefore the thermodynamic stability depends on $U_{CT}$. For example, in figure \ref{figs}, we see for $\log(U_s/U_{CT})=2$ that $F(M)$ is concave down indicating instability at $C=M=0$.

On the other hand, it is generally established that the thermodynamic
stability can be inferred from the presence or absence of poles of
quasi-normal modes in the upper half of the complex $\omega$ plane
\cite{Hartnoll:2009sz}. The quasi-normal mode analysis, however, does
not depend on holographic renormalization counter-term.  As such, it
would appear that thermodynamic stability is independent of
$U_{CT}$. But this is in direct contradiction with what we stated in
the previous paragraph.

The eventual resolution of this apparent tension can be understood as susceptibility being dependent on $U_{CT}$ but not the stability. But there are number of subtleties involved in arriving at this conclusion which we will describe in this subsection.

The issue boils down to mapping out the allowed range of values in
enumerating the renormalization schemes which $U_{CT}$ is
parameterizing. On the other hand, $\chi_\theta(M=0)$ depends on
$U_{CT}$ and as such can also be considered as parameterizing the
renormalizaiton schemes. The susceptibiltiy $\chi_\theta(M=0)$ however
is a quantity that is easy to measure. $U_{CT}$, on the other hand, is
an abstract quantity appearing in the counter-term which can only be
inferred by measuring some physical quantity (such as
$\chi_\theta(M=0)$) and using its relationship to $U_{CT}$. The choice
to parameterize the renormalization scheme with $\chi_\theta$ at $M=0$
is an arbitrary choice. Any other $M$ can be used as a reference. The
situation is analogous to the relation between renormalized coupling
in $\overline{MS}$ scheme and physical coupling inferred from
scattering at some definite energy. The former is the analogue of
$U_{CT}$ whereas the latter is the analogue of $\chi_\theta(M=0)$.

The question then is what constitutes the appropriate range of
parameters to enumerate distinct renormalization scheme. Should it be
$0 \le U_{CT} \le \infty$, or $-\infty \le \chi_\theta(M=0) \le
\infty$? To the extent that $\chi_\theta(M=0)$ is the physical
parameter, it would seem natural to treat the latter as parameterizing
physically distinct renormalization schemes. We will adopt that point
of view in this paper. There is a possibliy that extrapolation beyond
infinite $\chi_\theta(M=0)$ would admit interpretation along the lines
of dualities where one extrapolates beyond infinite coupling, but we
will not pursue that possiblity in this paper.

This implies however that the behavior illustrated in figure
\ref{figq} and \ref{figs} for $\log(U_s/U_{CT})=2$ where the
$\chi_\theta(M=0)$ is negative signalling instability corresponds to
pushing $\chi_\theta(M=0)$ beyond infinity and should be excluded from
our analysis. At $M=0$, one can explore the full range of $0 <
\chi_\theta(M=0)<\infty$ by letting $U_s/U_{CT}$ vary. Susceptibility
at $M=0$ for this system is always positive.

\begin{figure}
  \centerline{\includegraphics[width=4in]{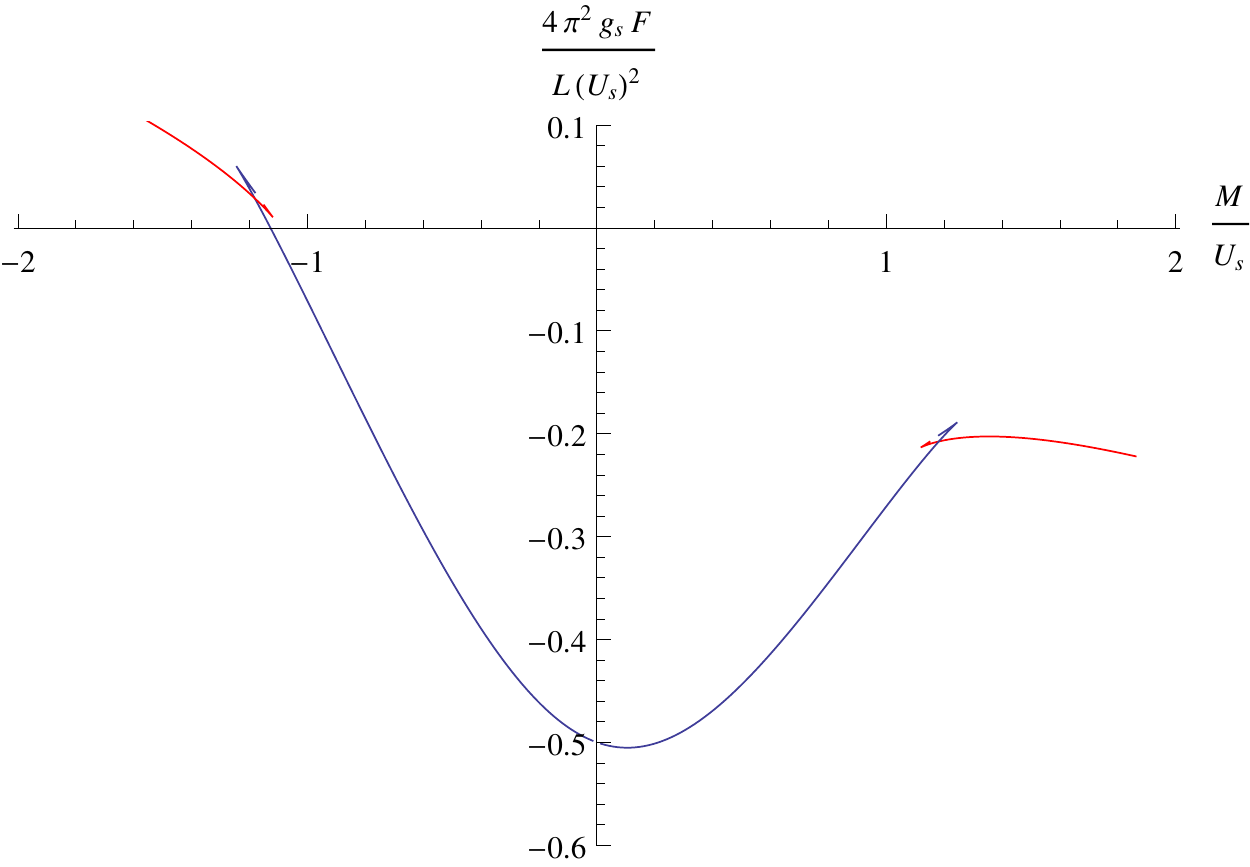}}
  \caption{$F_{eff}(M) = F(M) - C_{ext} M$ with $C/U_s=0.1$. \label{figt}}
\end{figure}

This however does not imply that the system is always stable or that
the stability analysis is completely unrelated to the quasi-normal
mode analysis. To see this, suppose we set the  temperature $U_s=U_{CT}$ so that the equation of state is given by  what is illustrated in figure \ref{figf}.  For every $0<C < C_{max}$, there is a subleading branch of solutions where 
\be {dC \over dM} < 0 \  . \ee
One can see the same thing by looking at unstable stationary point in the potential illustrated in figure \ref{figs} which is appropriately tilted by the inclusion of $-C_{ext} M$ term as is illustrated in fugre \ref{figt}.  One exepcts to find a corresponding pathology in the spectrum of quasi-normal modes for the fluctations around such unstable background solutions, but how can one understand its appearance short of doing an explicit computation?

One way to see that a cross-over into unstable behavior is taking
place is to focus on the state\footnote{Note that the position $C_{max}$ is independent of $U_{CT}$ since a change of $U_{CT}$ can only shift $\chi_{\theta}^{-1}$ by a finite amount.} at $C = C_{max}$ where  $\chi_\theta  = dC/dM=0$.  Recall that for every point on figure \ref{figf} there is a corresponding solution $\theta(U)$ which extremizes the action (\ref{nonlin}).  If we parameterize these solutions for fixed value of $M$ by $\theta_M(U)$, then at $M=M_{crit}$ corresponding to $C=C_{max}$, it follows that
\be \psi(U) = \left. {d \over dM} \theta_M(U)\right|_{M=M_{crit}}, \ee
$\psi(U)$ is a gap-less quasi-normal mode with $\omega=k=0$. The fact
that a gapless mode is appearing precisely when the susceptibility
$\chi_\theta=0$ strongly suggests that an unstable pole would appear
upon continuing to the branch where the susceptibility is
negative. This cross-over across $\chi_\theta=0$ is different from
crossin over $\chi_\theta=\infty$ which we discussed earlier in the
context of placing a bound on $U_{CT}$.

By a similar token, for sufficiently large negative value of
$\log(U_s/U_{CT})$, we encounter a point in the unstable branch where
$\chi_\theta$ reaches negative infinity. This can be seen in figure
\ref{figq} where the tangent of $M(C)$ is horizontal. For
$\log(U_s/U_{CT}) = -2$, this point corresponds to the switch back
point in the black hole embedding branch illustrated on the left most
figure of \ref{figs}.  This issue is somewhat academic, however, since
the switch back point is already subdominant in the saddle point
approximation as can be seen in figure \ref{figs}.

The critical point $(C,M) = (C_{max},M_{crit})$, on the other hand, is
a dominant saddle point and as such the resulting critical behavior
giving rise to a new channel for dissipating energy in the
hydrodynamics limit near that point is a real physical feature of this
model, at least at the perturbative level. This critical point can
also be seen to correspond to the point in figure \ref{figr} where
$T=T_{min}$ takes on a minimal value. The entropy $S$ plotted as a
function of $T$ in figure \ref{figr} is double valued. From the
thermodynamic point of view, the dominant branch, however, is
naturally the one with greater entropy. So, the black hole embedding
dominates, and the Minkowski embedding is the subdominant one. For $T
> T_{min}$, the system naively seems to be perfectly stable and well
behaved. At $T=T_{min}$, there is a critical behavior. The phase
strucure implied by these features are consistent with the $d=0$ slice
of the phase digram illustrated in figure 4 of
\cite{Hung:2009qk}. This however raises one obvious question. If at
$T=T_{min}$ we encounter a crticial behavior, where does the system
equilibriate to for $T < T_{min}$. In order to address this issue, we
need to go beyond the scope of perturbative stability analysis, and
consider the global issues. We will discuss that issue in the next
subsection.

The apparent quantitative mismatch in the dependence of $U_{CT}$ between susceptibility and quasi-normal mode specturm can also be seen in the computation of correlators in the real-time formalism. The retarded Greens function is computed using the prescription given in equation (3.15)  \cite{Son:2002sd}
\be G^R(k) = - 2 {\cal F}(k,z)|_{z_B} = \sqrt{-g} g^{zz} f_{-k}(z) \partial_z f_k(z) \ee
for a suitably normalized ingong wave $f(z)$, where $z \sim 1/u$. However, strictly speaking, ${\cal F}$ is divergent for our model and a counter-term is needed to render this expression finite.

One can understand the origin of this divergence as arising from computing the two ponit function of operator whose dimension is $\Delta$ so that at short distance, it scales as
\be G(k) = \langle {\cal O}(x) {\cal O}(0) \rangle \sim {1 \over x^{2 \Delta}} \ee
which then in momentum space takes the form 
\be G(k) \sim {1 \over k^{d-2 \Delta}} \ee
for large $k$. The issue arises when $d - 2 \Delta \le 0$ so that
this correlation funciton do not decay at large $k$. This is the case
in our example because $d-2 \Delta = 0$. So strictly speaking, one expects the two point function to scale for large $k$ as   \cite{Karch:2005ms}
\be G(k)  \sim \log(k^2/\mu^2) \ee
for some scale $\mu$. We are however interested in the small $k$ behavior when the system is at finite temperature.

The two point function should then admit a spectral decomposition
\be G(k) = \int ds {\rho(s) \over k^2 + s} \ee
where by power counting, we know that $\rho(s)$ must asymptote to a constant at large $s$. The integral over $s$, however, does not converge and must be regulated, for instance, by adding a term
\be G(k) = \int ds \left({\rho(s) \over k^2 + s} - {\rho(s) \over \Lambda^2 + s}\right) = a_0 + a_2 k^2 + a_4 k^4 + \ldots \ee
The term added is a contact term in that it is independent of $k$. It
only affects the $a_0$ term in the small momentum expansion of
$G(k)$. The scale $U_{CT}$ and $\mu$ arises from these considerations,
which does affect the two point function and therefore the
susceptibility, but does not affect the pole structure of
$G(k)$.\footnote{The issue of subtle contribution from contact terms was also discussed in (2.12)--(2.14) of \cite{Policastro:2002tn}.}. Nonetheless, both the susceptibility and the quasi-normal mode
spectrum knows when the system is perturbatively unstable, and
exhibits the appropriate symptoms.

\subsection{Non-perturbative stability of the D3/D3' system}

In this section, we will discuss the subject of how the unstable
states relaxes to the true equilibrium state.  This issue requires
consideration beyond the perturbative analysis, but the subject is not
an unfamiliar one. The same issue arises in the phase structure of
liquid-gas transitions in the van der Waals model. Let us see how that
applies to the D3/D3' system under consideration.

It is a fundamental fact of statistical mechanics that the set of
accessible states parameterized by the order parameters form a convex
set. A nice historical review of this basic notion can be found in
\cite{fisher}. If one is working at fixed temperature $T$, in a system
with a single order parameter $M$, the region bounded by the curve
$F(M)$ must be convex, or equivalently, $F''(M)>0$. But sometimes, as
is the case here, by working out the equation of states for what one
believes is the dominant thermodynamic configuration, one finds
regions where $F''(M)$ is negative, signalling an instability. For our system, this can be seen very explicitly in figure \ref{figs} where $F(M)$ is concave down for large values of $M$. Equivalently, one can see regions where $dC/dM$ is negative in the equation state illustrated in figure \ref{figf} and \ref{figq}. 

The standard procedure when this happens is to observe that the system can acheve a state with lower free energy for fixed $M$ by being in a hetrogeneous co-existence state. A co-existence of two state will give rise to a configuration with $(M,F(M))$ interpolating between the pair of states. As a result, the maximal extension of the space of states allowed by considering co-existence states is precisely the convexification of $F(M)$ achieved by supplementing $F(M)$ with ``ruled surfaces'' in the terminology of   \cite{fisher}.

\begin{figure}
\centerline{\includegraphics{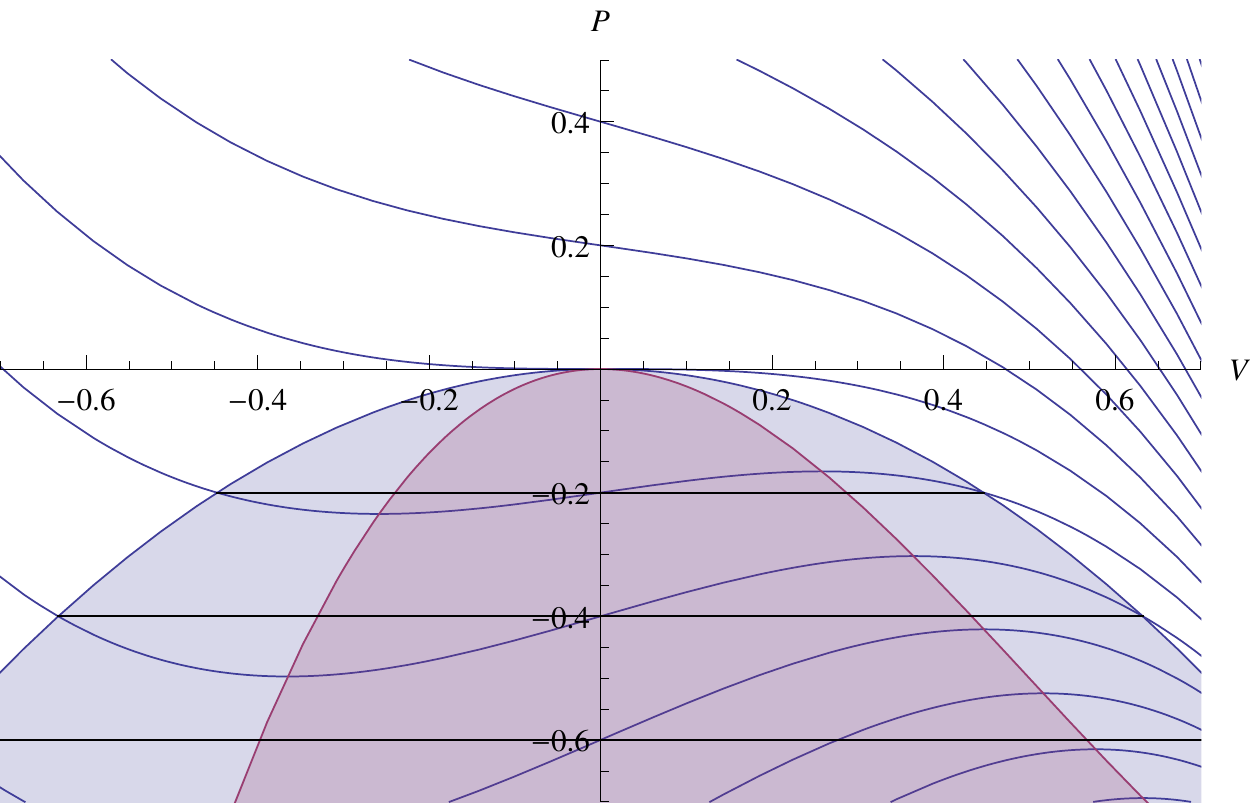}}
\caption{Typical equation of state $P(V)$ for various fixed $T$
  illustrating thermodynamic instability, coexistence of states, and
  the critical point. As $T$ is varied, regions of instability where
  $-\left.dP/dV\right|_T < 0$ appears and is shaded in red. This
  causes the system to undergo a first order phase transition where
  the equation of motion is modified by the ruled coexistence line,
  illustrated in black. The coexistence region is generally larger
  than the unstable region. So the local instability in the red shaded
  region affects the equilibrium configuration of all points in the
  $(P,V)$ region inside the blue shaded region.
\label{figg}}
\end{figure}

This is exactly what happens to the equation of state in van der Waals
system. We illustrate the standard diagram displaying a collection of
isothermal $P(V)$ curve in figure \ref{figg}. The point to note is the
fact that 1) there are regions where $-dP/dV <0$ signalling
instability, and 2) that this gives to a modified equation of state by
allowing coexistence states. It should also be emphasized that 3) the
region of phase diagram modified by allowing coexistence regions
(shaded blue region in figure \ref{figg}) are strictly greater than
the region where the system exhibts apparent instability (shaded red
region in figure \ref{figg}). 

How do these ideas apply to the D3/D3' system? From the equation of
state illustrated in figures \ref{figf} and \ref{figq} where $C(M)$ is
rapidly approaching zero as $M$ is increased, it follows that $F(M)$
must approach a constant value (set to zero for convenience in figure
\ref{figs}.) As soon as this potential is slighlty tilted in response
to non-vanishing external control parameter $C_{ext}$ as is
illustrated in figure \ref{figt}, the displaced local minima is no
longer a global minima and the system is susceptible to decay via
tunneling into a run away behavior towards large $M$. If $C_{ext}$ is
taken to be larger than $C_{max}$, even the local minima disappears
and the potential does not have any stationary points. (The situation
is a little different when the some net charge is introduced to the
D3'-brane world volume. We will ellaborate further on this case in a
separate publication.)

From the point of view of convexifying $F(M)$, we see that the result is to say
\be F(M) = F(M=0) = \mbox{constant} \ee
which is even more susceptible to runaway behavior when $C$ is non-vanishing.

What appears to be happening is that the $d=0$ slice of the phase diagram illustrated in figure 4 of \cite{Hung:2009qk} degenerated completely and that at finite $C$, the system is completely unstable at the non-perturbative level.

What this means, presumably, is that the classical treatment is
predicting its own thermodynamic demise and that some configuration
not presently accounted for will modify the equation of state to
provide the stable, equilibirium state for this system. Such
corrections, however, must arise from string or quantum corrections
and is expected to scale non-trivially with respect to $g_S$ and
$N$. It is possible that such correction would also give rise to the
branch drawn with dotted lines in figure 3.d of \cite{Hung:2009qk}.

Alternatively, the D3/D3' system is intrinsically unstable. At this moment, we do not have any reason to rule out that possiblity.

In this article, we took the boundary of $AdS_5 \times S_5$ to be flat
and infinite in volume. It would be interesting to repeat this
analysis treating the boundary to be a $S_3$ of finite size. In that
case, the entire system undergoes a Hawking-Page transition
\cite{Witten:1998zw}, giving rise to a qualitatively new behavior also
for the D3'-brane embedding. It is somewhat unexpected, however, for
thermodynamic stability of a system to rely on finite volume issue.
In any case, it would be very interesting to understand all the
different ways in which the run-away behavior seen in this system can
be stablized.

\section{Discussions}

In this article, we analyzed the embeddings of D3'-brane probe in
Schwarzschild $AdS_5$ geometry and studied their thermodynamic
interpretations, with emphasis on order parameter and control
parameter dual to the brane probe embedding psuedo scalar field
$\theta(U,x)$. We described the subtle relationship between the
anomalous scale $U_{CT}$ introduced in the holographic renormalizatoin
procedure, thermal susceptibitliy $\chi_\theta$, and the spectrum of
quasi-normal modes.

The resulting analysis of the theromdynamic stability revealed that
while the system is perturbatively stable for some range of
parameters, it is unstable at the non-perturbative level almost
everywhere. The full implication of this instability is not completely
clear to us at the moment. It should be noted that our consideration
was limited to treating the branes in the probe approximation. Perhaps
gravitational back reaction will stabilize the system, although properly 
addressing this issue is a tall order for an intersecting brane
system. It is also possible that a satisfactory resolution will
require going beyond the semi-classical treatment of these systems.

The original goal of this study was to find some relation between
subtle features of the thermodynamics of D3/D3' brane system to the
subtle features discussed recently by Mintun, Polchinski, and Sun
\cite{Mintun:2014aka}. At the moment, the only connection we see is
the fact that some of the pathologies are due to low co-dimension
physics in both instances. We hope to provide more insight into this
issue in the future.

Finally, let us comment that the intersecting D3/D3' system could also
prove to be useful as a probe of the region behind the horizon in the
context of entanglement entropy where the degrees of freedom crossing
the horizon is in the open string sector \cite{Maldacena:2001kr}. Just like in other attempts to probe behind the horizon such as \cite{Kraus:2002iv,Fidkowski:2003nf}, we expect the sensitivity of probe dyanmics to behind the horizon physics to be highly suppressed. Nonetheless, perhaps something can be gained by using an open string probe instead of a closed string probe.

\section*{Acknowledgements}

We are grateful to 
A.~Buchel, 
S.~Coppersmith, 
S.~Gubser,
V.~Hubeny,
M.~Kruczenski,
D.~Mateos, and 
W.~Taylor
for discssions and M.~Pillai for collaboration during the early stage
of this work. We also thank the referee of {\it Physical Review} {\bf
  D} for constructive review which dramatically improved the scope and
the content of this manuscript.

\appendix

\section{More on embeddings}

In this appendix, we offer an alternative parameterization of the equation of motion (\ref{nonlin}) where we map the problem to dynamics of a particle rolling down a potential. This formulation is useful for visualizing the solutions and for providing assurance that all solutions are accounted.

The procedure to convert (\ref{nonlin}) into the classical potential problem form is essentially the same steps one takes to convert the Nambu-Goto action in to the Polyakov action form.  This can be done by introducing an auxiliary world volume parameter $\tau$ and a
Lagrange multiplier $\lambda(\tau)$ and re-writing the effectively one
dimensional form of (\ref{nonlin}) as
\be \int d\tau \  {1 \over 2}   U(\tau) \cos (\theta(\tau))  \left[ \lambda(\tau)^{-1} (U'(\tau)^2 + (U(\tau)^2 - U_s^4 U(\tau)^{-2}) \theta'(\tau)^2) + \lambda(\tau) \right] \ . \ee
Solving for the Lagrange multiplier and setting $\tau=U$ recovers (\ref{nonlin}). On the other hand, imposing as the gauge condition 
\be \lambda(\tau) = {U^2 - U_s^4 U^{-2} \over U_s^2} U \cos \theta \ , \ee
setting
\be U^2 = U_s^2 \cosh s  \ , \ee
and rescaling $\tau = U_s \sigma$  will scale out $U_s$, leads to 
\be L = U_s^2 \int d \sigma \ {1 \over 2} \dot s^2 + {1 \over 2} s^2 \dot \theta^2 + {1 \over 2} \sinh^2 s \cos^2 \theta \label{s-theta}\ee
where $s$, $\theta$, and $\sigma$ are dimensionless, and dot denotes derivative with respect to $\sigma$. In the form (\ref{s-theta}) the problem is essentially that of a particle whose positions are parameterized by $1 < s < \infty$ and $-\pi/2 < \theta < \pi/2$ rolling along a potential 
\be V(s,\theta) =  - {1 \over 2} \sinh^2 s \cos^2 \theta \label{stV} \ . \ee
Reparameterization invariance further implies that the solution should correspond to trajectory with vanishing Hamiltonian 
\be {\cal H}=0 \ . \ee

\begin{figure}
\centerline{\includegraphics[width=4in]{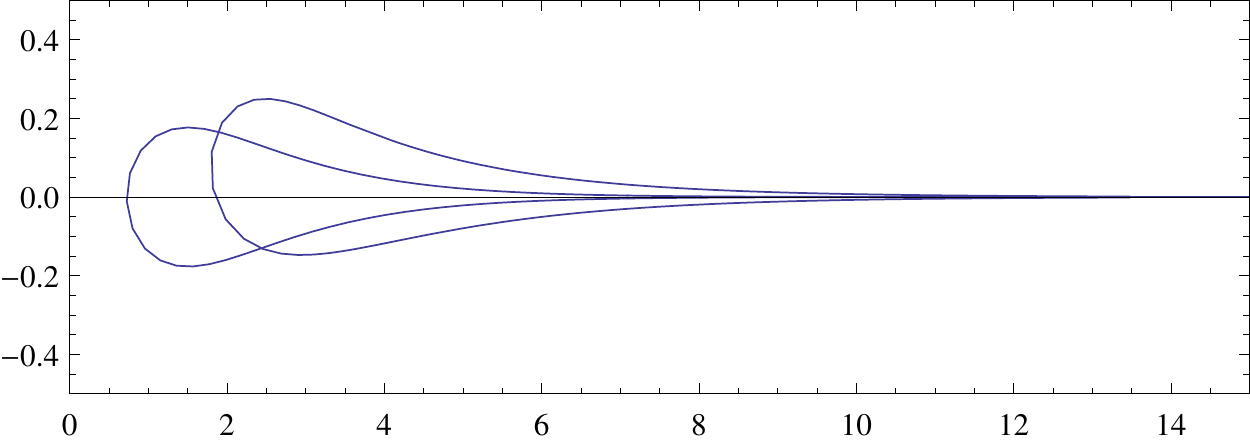}}
\caption{The trajectory of figure \ref{figb} in $(s,\theta)$ plane. The horizontal axis is $\log(s)$ and the vertical axis is $\theta/\pi$. The horizon corresponds to the left edge at  $\log(s)=0$.  \label{figc}}
\centerline{\includegraphics[width=3in]{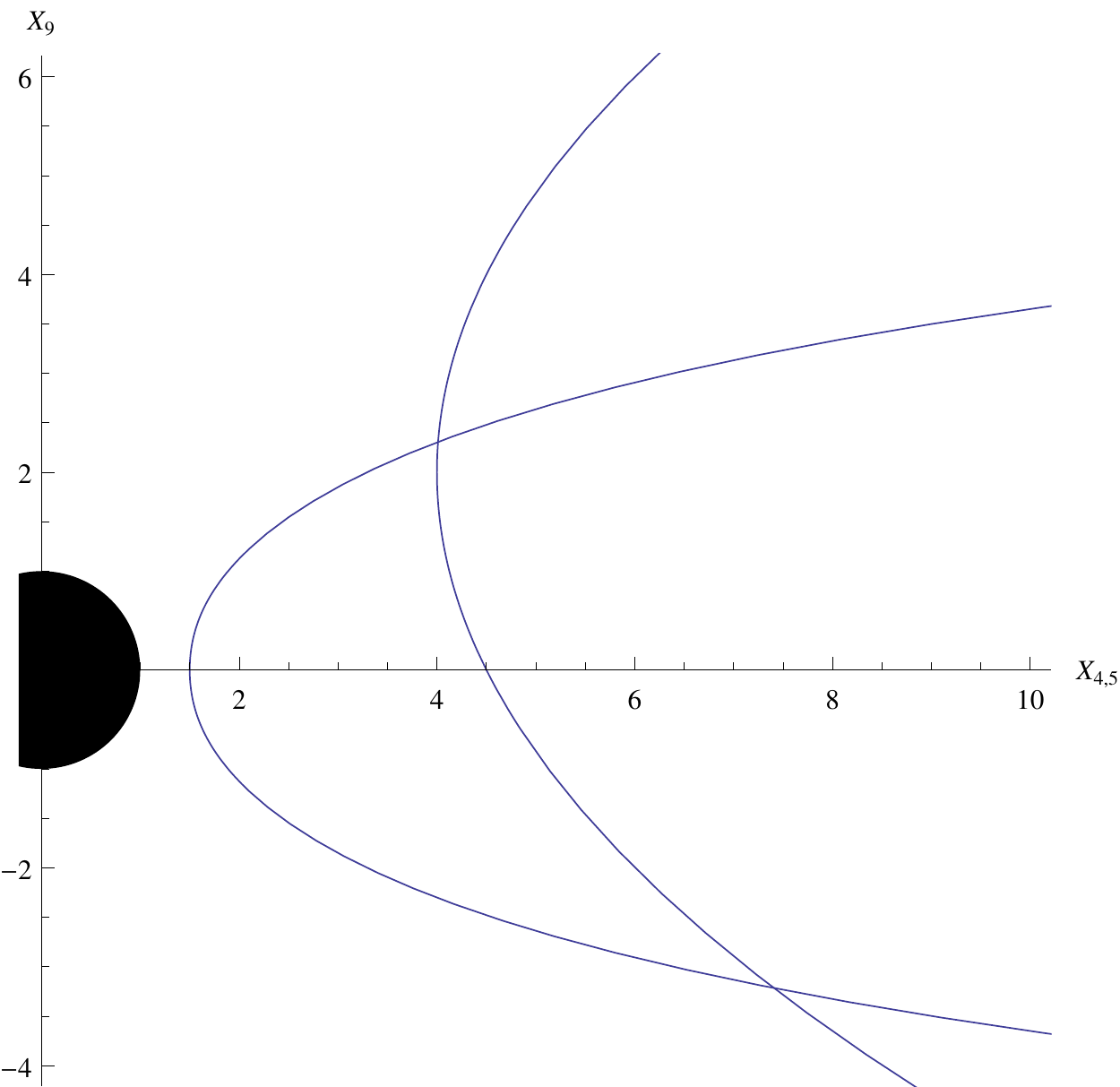}}
\caption{Some generic solutions to the embedding equation of a D3'-brane in a finite temperature $AdS_5 \times S_5$ background. The full embedding is cylindrically symmetric with respect to rotation around the $X_9$ axis. The black disk represents the region behind the event horizon.\label{figb}}
\end{figure}

A large class of solutions arises as a trajectory of 
 a particle rolling up, turning around, and rolling back
down the potential in $(s,\theta)$ coordinates (\ref{stV}) which we
illustrate in figure \ref{figc}. These are the worm hole embeddings in the terminology of \cite{Christensen:1998hg,Frolov:1998td}. In the original $(X_{4,5},X_9)$ cooridinates, they look like an embedding illustrated in figure \ref{figb}. In the large $X$ region, these correspond to having both a brane and an anti-brane along the lines of \cite{Callan:1997kz,Sakai:2004cn} and are not the solutions we are looking for.

The only other possibility is for the trajectory in the $(s,\theta)$
plane to hit the boundary of the region on which the space is defined,
i.e. $s=1$ for arbitrary $\theta$, or $\theta = \pi/2$ for arbitrary
$s$.  These solutions are referred to as the black hole and the
Minkowski embeddings, respectively. These trajectories are specified
uniquely by giving the starting position of the trajectory along
the boundary of the $(s,\theta)$ plane because the quadratic term in
the equation of motion inferred from (\ref{nonlin}) degenerates there.

\begin{figure}
\centerline{\includegraphics[width=4in]{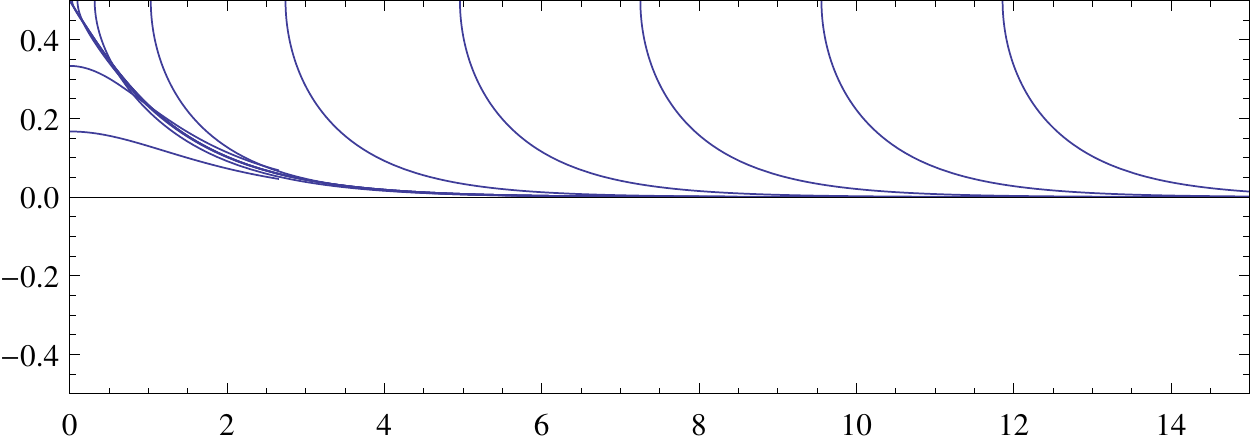}}
\caption{The trajectories of figure \ref{figd} in the $(s,\theta)$ plane. The horizontal axis is $\log(s)$ and the vertical axis is $\theta/\pi$. \label{fige}}
\end{figure}

The trajectory resulting from these initial conditions are illustrated
in figure \ref{fige}. The same solutions in the original
$(X_{4,5},X_9)$ coordinates is illustrated in figure \ref{figd}. The
trajectories starting at $s=1$ boundary are the black hole embedding,
and the trajectories starting at $\theta = \pi/2$ are the Minkowski embedings.

\section{Retarded Green Function and Quasi-Normal Mode for $\theta(U)=0$ background\label{appB}}

In this appendix, we will outline the computation of retarted Green function for the $\theta(U)$ fluctuation using the prescription of \cite{Son:2002sd}. We start by generalizing (\ref{nonlin}) to include momentum and energy and write
\be S_{DBI}  = {1 \over (2 \pi)^2 g_s} \int d^2 x \, dU \, U     \cos (\theta(U)) \sqrt{\left(1+ f U^2\right)  \theta'(U)^2+ {\lambda (f k^2 - \omega^2) \over U^2} \theta(U)^2}  \label{qn}  \ . \ee
In order to compute the retarded Green function, we need to expand $\theta(U)$ to quadratic order around a solution to the equation of motion. This is quite complicated for generic solution $\theta_0(U)$, but is tractable for the trivial solution $\theta_0(U) = 0$.  For that case, the linearized action reads
\be S_{lin} \sim  {1 \over 2} U \left( U^2 f \theta'(U)^2 - \theta(U)^2 + {\lambda (f k^2 - \omega^2) \over f U^2} \theta(U)^2 \right) \ee
The equation of motion resulting from this action can be written in a canonical form in terms of
\be \theta(x) = (x-1)^{1/2} y(x) \ , \qquad x = {U^2-U_s^2 \over U^2 + U_s^2} \, \ee
with
\be \omega = {U_s w \over \sqrt{\lambda}}, \qquad k = {U_s q \over \sqrt{\lambda}} \ee
so that the equation of interest is expressed as
\beq 0 & = &  y''(x) - \left({1 \over 1-x} - {1-i w/2 \over x} \right) y'(x) \cr
&& 
+\left( \frac{k^2-w^2+2}{8 (x-1)}+\frac{-2 q^2+w^2-2}{8 x}+\frac{k^2}{8
   (x+1)}+\frac{w^2}{16 x^2}\right)  y(x) \ . 
\eeq
For non-zero $q^2$, this equation has four singular points at $x=0$,
$x=1$, $x=-1$, and $x=\infty$, and as such is not analytically
tractable. But when $q^2=0$, the singularity at $x=-1$ goes away, and
one arrives at a hypergeometric equation solved by
\beq y(x) &=& c_1 (-1)^{-\frac{i w}{4}} x^{-\frac{i w}{4}} \,
   _2F_1\left(\frac{1}{2}-\left(\frac{1}{4}+\frac{i}{4}\right)
   w,\left(\frac{1}{4}-\frac{i}{4}\right) w+\frac{1}{2};1-\frac{i
     w}{2};x\right) \cr
   && + c_2 (-1)^{\frac{i w}{4}} x^{\frac{i w}{4}} \,
   _2F_1\left(\frac{1}{2}-\left(\frac{1}{4}-\frac{i}{4}\right)
   w,\left(\frac{1}{4}+\frac{i}{4}\right) w+\frac{1}{2};\frac{i
     w}{2}+1;x\right) \ . 
\eeq
One of the solution satisfies the infalling boundary condition at the horizon and the other is outgoing. We can then find the large $U$ asymptotics the ingoing solution and find it to be of the form
\be \theta(U) = A \log(U/U_s) + B \ee
where
\beq A & = & \frac{2 \sqrt{2}  e^{\frac{\pi  w}{4}} \Gamma
   \left(1-\frac{i w}{2}\right)}{\Gamma
   \left(\frac{1}{2}-\left(\frac{1}{4}+\frac{i}{4}\right) w\right)
   \Gamma \left(\left(\frac{1}{4}-\frac{i}{4}\right)
   w+\frac{1}{2}\right)} U_s \\
B & = & -\frac{\sqrt{2} e^{\frac{\pi  w}{4}} \Gamma \left(1-\frac{i
   w}{2}\right) \left(\psi
   ^{(0)}\left(\frac{1}{2}-\left(\frac{1}{4}+\frac{i}{4}\right)
   w\right)+\psi ^{(0)}\left(\left(\frac{1}{4}-\frac{i}{4}\right)
   w+\frac{1}{2}\right)+2 \gamma +\log (2)\right)}{\Gamma
   \left(\frac{1}{2}-\left(\frac{1}{4}+\frac{i}{4}\right) w\right)
   \Gamma \left(\left(\frac{1}{4}-\frac{i}{4}\right)
   w+\frac{1}{2}\right)} U_s \nonumber \eeq
from which we infer that
\be G_R(w) = {B\over A} -  \log(U_s/U_{CT}) \label{OOret} \ee
with
\be {B \over A} = \frac{1}{2} \left(-\psi
   ^{(0)}\left(\frac{1}{2}-\left(\frac{1}{4}+\frac{i}{4}\right)
   w\right)-\psi ^{(0)}\left(\left(\frac{1}{4}-\frac{i}{4}\right)
   w+\frac{1}{2}\right)-2 \gamma -\log (2)\right) \  . \ee
We see that the poles of the retarded Green function is encoded in the poles of $\psi^0(x)$ function and is independent of $U_{CT}$, but the Green function itself is dependent on $U_{CT}$ through momentum independent contact terms as is shown in (\ref{OOret}).

\section{Holographic dictionary and the effective action for the order parameter}

In this article, the holographic dictionary (\ref{AdSdict0}) and (\ref{AdSdict})
\be \langle e^{\int d^d C(x) {\cal O}(x)} \rangle_{boundary} = Z_{bulk}[C(x)] =  \int [D \theta(U,x)]_{C(x)} e^{-(S_{DBI}[\theta(U,x)]+S_{CT}[\theta(U,x)])} \label{AdSdict2} \ee
played a critical role in providing an interpretation of brane embeddings in the bulk of space time in terms of field theory observables. The dependence on control parameter/boundary condition $C(x)$ is somewhat implicit in the path integral expression on the right most side of (\ref{AdSdict2}). In this appendix, we will provide a formal path integral manipulation to make this explicit, as well as derive a formal path integral expression for the effective action for the field of the  expectation value of the operator $M(x) \sim \langle {\cal O}(x) \rangle$ corresponding to the bulk field $\theta(U,x)$.

The trick is to introduce an $M(x)$ as an auxiliary field which when integrated out reporduces the original path integral as follows.
\be Z_{bulk} [C(x)]  =  \int [D\theta(U,x)][D M(x)] e^{-(S[\theta(U)] +  {1 \over(2 \pi)^3 g_s} \int d^2x \left( {U_{UV}\theta(U_{UV},x)\over \log(U_{UV}/U_{**})} M(x) - C(x) M(x)\right))}  \label{AdSdict3}
\ee
where we are working in the path integral with cutoff at $U=U_{UV}$ to
regularize the contribution of the counter-term although we will take
the $U_{UV} \rightarrow \infty$ limit in the end. The auxiliary field
lives only at $U=U_{UV}$. The presence of the holographic
counter-terms guarantees that this limit is smooth. Because a
logarithm is involved, we have intrduced yet another scale $U_{**}$
although the dependence on this scale drops out in the $U_{UV} \rightarrow \infty$ limit. It is clear that integrating out $M(x)$ will impose the boundary condition and reproduces (\ref{AdSdict2}).

The expression (\ref{AdSdict3}) is useful for a variety of
reasons. First, note that functionally differentiating with respect to
$C(x)$ pulls down a $M(x)$. In that sense, we immediately associate
$M(x)$ with the expectation value $\langle {\cal O}(x) \rangle$.

We can also read off a path integral expression for the effective action of $M(x)$ easily as follows.
\be e^{-{1 \over(2 \pi)^2 g_s} \Gamma[M(x)]}  =  \int [D\theta(U,x)] e^{-(S[\theta(U)] + {1 \over(2 \pi)^2 g_s} \int d^2x \left({U_{UV}\theta(U_{UV},x)\over \log(U_{UV}/U_{**})} M(x)\right))}\ .   \label{AdSdict4}
\ee
In this expression, the term linear in $M(x)$ is imposing a boundary
condition for the path integral over $\theta(U,x)$. It is also clear
that $\Gamma[M(x)]$ and $W[C(x)]=- \log(Z[C(x)])$ are related by the
standard Legendre transform at the leading order in saddle point
approximation, whose correction can be computed systematically \cite{Klebanov:1999tb}.

The effective action $\Gamma[M(x)]$ is a complicated expression which
includes terms with arbitrary orders of $M(x)$ and its
derivatives. But it is formally defined umambiguously in
(\ref{AdSdict4}) and can be computed systematically as an expansion in
$M(x)$ and its derivatives.

\bibliography{embedBH}\bibliographystyle{utphys}

\end{document}